\documentclass[sort&compress]{elsarticle}

\usepackage{lineno,hyperref}
\modulolinenumbers[5]
\usepackage{epsfig} %% for loading postscript figures
\usepackage{graphicx}
\usepackage{caption}
\usepackage{subcaption}
\usepackage{multirow}
\usepackage{multicol}
\usepackage{etoolbox}
\usepackage{setspace}
\usepackage{lipsum}
\makeatletter
\def\ps@pprintTitle{%
 \let\@oddhead\@empty
 \let\@evenhead\@empty
 \def\@oddfoot{}%
 \let\@evenfoot\@oddfoot}
\makeatother
\usepackage{amssymb,mathtools}
\usepackage{lscape}

\usepackage[utf8]{inputenc}

\usepackage{siunitx}
\usepackage[nocfg]{nomencl}
\usepackage{nomencl}
\usepackage{ifthen}
\renewcommand\nomgroup[1]{%
  \ifthenelse{\equal{#1}{A}}{%
    \item[\textbf{Acronyms}]}{%                A - Acronyms
  \ifthenelse{\equal{#1}{R}}{%
    \item[\textbf{Roman Symbols}]}{%           R - Roman
  \ifthenelse{\equal{#1}{G}}{%
    \item[\textbf{Greek Symbols}]}{%           G - Greek
  \ifthenelse{\equal{#1}{S}}{%
    \item[\textbf{Superscripts}]}{%            S - Superscripts
  \ifthenelse{\equal{#1}{U}}{%
    \item[\textbf{Subscripts}]}{%              U - Subscripts
  \ifthenelse{\equal{#1}{X}}{%
    \item[\textbf{Other Symbols}]}{%           X - Other Symbols
  {}}}}}}}}

\makenomenclature

\usepackage{fancyhdr}
\date{}
\begin{document}

\title{Investigations of Non-Gray/Gray Radiative Heat Transfer  Effect on Natural Convection in Tall Cavities at Low Operating Temperature}

%\author{G. Chanakya and Pradeep Kumar}
\author[]{Pradeep Kumar, G. Chanakya  and Naman Barthwal}
\author[]{\thanks{corresponding author}}

\address{Numerical Experiment Laboratory \\ 
	 (Radiation and Fluid Flow Physics)\\
	School of Engineering\\
	Indian Institute of Technology Mandi\\
	Mandi, Himachal Pradesh, India 175075\\}

\ead{pradeepkumar@iitmandi.ac.in}

%\maketitle
\begin{frontmatter}
\begin{abstract}
The present work numerically investigates the influence of radiation on the natural convection in differentially heated slender and square-base tall cavities at low operating temperature range of 288.1-307.7 K. The atmospheric air which has the composition of $N_{2}$, $O_{2}$, $CO_{2}$ and $H_{2}O$ in the molar mass  proportion of 75.964\%, 21\%, 0.036\% and 3\%, respectively, has been considered as the working fluid in the cavities. The non-gray and gray Planck mean absorption coefficients for whole spectrum of the air at the atmospheric pressure and average temperature of the cavity have been calculated from HITEMP-2010 database by Line-By-Line (LBL) approach. The flow and heat transfer simulations for combined radiation and natural convection have been performed for four different scenarios, i.e., pure convection, radiation in transparent, gray, and non-gray radiation medium. First, the numerical results are compared with the experimental results of Betts and Bukhari \cite{betts2000experiments} for the slender cavity case, subsequently a comprehensive analysis of the fluid flow and heat transfer are presented to demonstrate the influence of the radiation on natural convection inside both the cavities. The flow structure has been visualized by Q-criterion - a vortex identification technique. The results reveal that the fluid flow and thermal characteristics change significantly at the top and  the bottom of both the cavities with the inclusion of radiation. Moreover, the radiation in gray medium has significant effect on these characteristic. The fluid flow and heat transfer are only happening in a narrow regions near to the active walls. The radiative flux is in the same order of the conductive flux in these cavities and these fluxes are almost same in all four scenarios of the radiation modeling.

\end{abstract}
%end{frontmatter}
%\section*{keywords}

\begin{keyword}
\texttt{Non-Gray Radiation, Natural Convection, Tall Cavities, HITEMP Database, Line-by-line Method.}
%\MSC[2010] 00-01\sep  99-00
\end{keyword}

\end{frontmatter}

%\linenumbers

\nomenclature[a]{$a_{\lambda} =$}{Spectral absorption coefficient, ~~ $\frac{1}{m}$}
\nomenclature[a]{$c_{p} =$}{Specific heat, ~~ $\frac{J}{kg-K}$} 
\nomenclature[a]{$D =$}{Characteristic length, depth, $m$}
\nomenclature[a]{$D_{\omega} =$}{Cross-diffusion term, ~~ $\frac{kg}{m^3 s^2}$}
\nomenclature[a]{$F =$}{Fractional Planck function}
\nomenclature[X]{$\tilde{G}_{\omega} =$}{ Generation term for $\omega$,~~$\frac{kg}{m^3 s^2}$}
\nomenclature[a]{$H =$}{Height,~~$m$}
\nomenclature[X]{$I_{\lambda} =$}{Spectral intensity,~~$\frac{W}{m sr}$}
\nomenclature[a]{$k =$}{Conductivity,~~$\frac{W}{m K}$, Turbulent kinetic energy,~~$\frac{m^2}{s^2}$}
\nomenclature[X]{$\hat{n} =$}{Unit area normal vector}
\nomenclature[a]{$Nu_{C} =$}{Conductive Nusselt number}
\nomenclature[a]{$Nu_{R} =$}{Radiative Nusselt number}
\nomenclature[a]{$p =$}{Pressure,~~$\frac{kg}{m s^2}$}
\nomenclature[a]{$Pr =$}{Prandtl number}
\nomenclature[a]{$q_{C} =$}{Conductive heat flux,~~$\frac{W}{m^2}$}
\nomenclature[a]{$q_{R} =$}{Radiative heat flux,~~$\frac{W}{m^2}$} 
\nomenclature[a]{$Ra =$}{Rayleigh number, $\frac{g\beta(T_h-T_c)D^3}{\nu\alpha}$}
\nomenclature[X]{$\vec{r} =$}{Position vector}
\nomenclature[X]{$\vec{s} =$}{Direction vector}
\nomenclature[a]{$T =$}{Temperature,~~$K$}
\nomenclature[a]{$u =$}{Velocity,~~$\frac{m}{s}$}
\nomenclature[a]{$W =$}{Width,~~$m$}
\nomenclature[a]{$Y_{k} =$}{Dissipation term for $k$,~~$\frac{kg}{m s^3}$}
\nomenclature[a]{$Y_{\omega} =$}{Dissipation term for $\omega$,~~$\frac{kg}{m^3 s^2}$} 
%\textit{Greek Symbols}\\
\noindent\nomenclature[g]{$\delta_{ij} =$}{Kronecker delta}
\nomenclature[g]{$\mu =$}{Dynamic viscosity,~~$\frac{kg}{m s}$}
\nomenclature[g]{$\nu =$}{Kinematic viscosity,~~$\frac{m^2}{s}$}
\nomenclature[g]{$\alpha =$}{Thermal diffusivity,~~$\frac{m^2}{s}$}
\nomenclature[g]{$\phi =$}{Scalar}
\nomenclature[g]{$\theta =$}{Polar angle}
\nomenclature[g]{$\Phi =$}{Azimuth angle}
\nomenclature[g]{$\Gamma =$}{Effective diffusivity,~~$\frac{m^2}{s}$}
\nomenclature[g]{$\mu_{t} =$}{Turbulent viscosity,~~$\frac{m^2}{s}$}
\nomenclature[g]{$\lambda =$}{Wavelength,~~$m$}
\nomenclature[g]{$\omega =$}{Specific dissipation rate,~~$\frac{1}{s}$}
\nomenclature[g]{$\Omega =$}{Solid angle,~~$sr$}
\nomenclature[g]{$\rho =$}{Density,~~$\frac{kg}{m^3}$}
\nomenclature[g]{$\epsilon_{\lambda, w} =$}{Spectral wall emissivity}
\nomenclature[g]{$\sigma =$}{Stefan-Boltzmann constant,~~$\frac{W}{m^2 K^4}$}
%\textit{subscripts}\\
\nomenclature[U]{$i,j =$}{Tensor indices}
\nomenclature[U]{$ref =$}{Reference point}
\nomenclature[U]{$R =$}{Radiation quantity}
\nomenclature[U]{$C =$}{Conduction quantity}
\nomenclature[U]{$t =$}{Turbulent quantity}
\nomenclature[U]{$b =$}{Black body quantity}
\nomenclature[U]{$\lambda =$}{Spectral quantity}
\nomenclature[U]{$T =$}{Total quantity}
%\textit{Acronyms}\\
\noindent
\nomenclature[a]{CDSD =}{Carbon dioxide spectroscopic databank}
\nomenclature[a]{CHF =}{Convductive heat flux}
\nomenclature[a]{HITEMP =}{The high-temperature  molecular spectroscopic database}
\nomenclature[a]{HITRAN =}{High-resolution transmission molecular absorption database}
\nomenclature[a]{LBL =}{Line-By-Line}
\nomenclature[a]{RHF =}{Radiative heat flux}

\printnomenclature

\section{Introduction}
The fluid flow and the heat transfer play very important role in every walk of life ranging from human comfort to the high density energy applications such as in refrigeration/air conditioning, building energy analysis, combustion, nuclear reactors, etc. Hence, it becomes imperative to study and analyze these phenomena in details. Generally, in any situation of heat transfer, all modes of heat transfer are always present, however, it is very natural to ignore the radiative heat transfer in the most of the cases. The radiative heat transfer complicates analysis due to its directional nature and variation of properties with pressure, temperature and wavelength. In order to understand the realistic effect of the radiation in a participating medium, it is necessary to know the spectral radiative properties such as absorption/emission, scattering coefficients. These properties are highly dependent on the pressure, temperature and wavelength. A bulk of literature is available on the heat transfer in natural \cite{kumar2013effect, de1983natural, fusegi1991numerical, byun2002radiation} forced \cite{kumar2008effect, atashafrooz2012combined, bouali2006combined, azad1981combined, tabanfar1987combined, ansari2010thermal, ansari2011study, ansari2013forced} and mixed convection \cite {chiu2008mixed}, where radiation heat transfer is either neglected or included by assuming either transparent or gray radiation medium. Only a few literature have considered the radiation heat transfer with spectral properties of medium in the analysis of the natural convection\cite{lari2012numerical,soucasse2014transitional,soucasse2012numerical,ibrahim2013coupling}. 

The experimental works of Tsuji and Nagano \cite{tsuji1988turbulence} for a heated vertical plate, by Betts and Bokhari \cite{betts2000experiments} for a tall cavity, by King \cite{king1989turbulent} and Cheeswright et al. \cite{cheesewright1986experimental} for a rectangular cavity, by Tian and Karayiannis \cite{tian2000low} and Ampofo and Karayiannis \cite{ampofo2003experimental} for a square cavity have been referred quite frequently for bench marking of the solutions by many researchers. Wu and Lei \cite{wu2015numerical} reported the discrepancy in interior stratification between the experimental and the numerical simulation which was mainly caused by the negligence of the radiation heat transfer and the results in the interior of the cavity have improved with the consideration of the radiation model. It was also found that a slightly better prediction of the temperature profile was achieved when the study was extended to three-dimensional simulation and results were better agreement with the experimental data.  Xin et al. \cite{xin2013resolving} used the direct numerical simulation (DNS) approach for a three-dimensional differentially heated cavity and concluded that the surface radiation was an important factor that affected the natural convection in air-filled cavities, thus should not be neglected. Akiyama and Chong \cite{akiyama1997numerical} conducted numerical analysis of natural convection with the surface radiation and observed that the surface radiation altered the distribution of the temperature field on the passive walls. The temperature gradients near the active walls were relatively small, and the thermal boundary layer was thickened by surface radiation. Ibrahim et al. \cite{ibrahim2013coupling} investigated the effect of the radiation on the natural convection in a two-dimensional differentially heated square cavity. The flow structure inside the cavity did not get much affected by considering the gas radiation alone and also by alone considering the wall radiation. Much more pronounced effects were observed when both the wall and the gas radiation were considered together which also resulted in increased turbulence levels. Kumar and Easwaran \cite{kumar2010numerical} did the combined radiation and natural convection simulation of a three-dimensional cubic cavity. They highlighted that flow became more and more organized with increased optical thickness and also the radiative heat transfer dominated the conductive heat transfer on the walls. The temperature field became two-dimensional with increased optical thickness of the medium, whereas the flow field was still three-dimensional. A thorough work on the convective flow with radiation in canonical problems can be found, for example, in the review papers by Ostrach \cite{ostrach1988natural} and Kumar \cite{kumar2009,kumar2015,kumar2013effect}, in the books by Martynenko and Khramtsov \cite{martynenko2005free}, and Bejan \cite{bejan2013convection}. Mostly the gray radiative properties have been used in all these works discussed above. 
 
 The review of above literature indicates that radiation is an important phenomenon in natural convection and has been modeled in many ways, but the realistic modeling of radiation requires non-gray radiative properties. In a bid to calculate the non-gray radiative properties led to the development of various databases \cite{modest2013radiative} such as HITRAN (high-resolution transmission molecular database) \cite{rothman2009hitran}, HITEMP (the high-temperature molecular spectroscopic database) \cite{rothman2010hitemp}, CDSD (carbon-dioxide spectroscopic databank) \cite{tashkun2011cdsd}, etc. These databases provide the line transition on a wavenumber for different chemical species over full spectrum of thermal radiation. HITEMP database contains the line transition information at 296 K and 1 atmospheric pressure. It further requires to perform mathematical modeling and computation to obtain the desired properties at different temperature, pressure and mole fraction from these databases. The transmissivity and the absorptivity calculations for chemical species like $H_{2}O$, $CO_{2}$ etc in isothermal and homogeneous medium have been performed by Alberti et al. \cite{alberti2015validation} and Chu et al. \cite{chu2016effects}. Some experimental results for the transmissivity and the absorptivity of these chemical species are also available \cite{chu2016effects,soufiani1997high} and have been extensively used for the benchmarking purposes of the theoretical calculations. Bartwal et al. \cite{bartwal17bits, bartwal17iitm} have also performed these calculations to obtain absorption coefficient of single specie, such as $CO_{2}$, $H_{2}O$, mixture of gases and air \cite{bartwal18china} using the HITEMP database in a homogeneous and isothermal medium.

The present work deals with the investigation of non-gray/gray modelings of the radiative heat transfer in natural convection problem in two three-dimensional tall cavities. This is performed by making a comparison study among pure convection (without radiation), radiation in non-participating, gray, and non-gray medium. The atmospheric air which has composition of $N_{2}$, $O_{2}$, $H_{2}O$, and $CO_{2}$ in molar mass of 75.964\%, 21\%, 0.036\% and 3\%, respectively, is considered as the working fluid. Other gases (like methane, nitrous oxide etc) are present in atmosphere in very small percentages, thus they are neglected in the radiative properties calculations. The non-gray radiative property of air has been calculated by line-by-line method using HITEMP-2010 database, whereupon Planck mean of spectral absorption coefficient over whole spectrum has been calculated and used as a gray absorption coefficient of the air. 

The outline of present manuscript is as follows: section 2 describes the problem definition, general assumptions, mathematical modeling, boundary conditions, followed by section 3 for sensitivity in selection of grids, direction and spectral bands; the validation work is presented in section 4, section 5 presents results and discussion, and finally concluding remarks is in section 6.

\section{Problem Statement}
\begin{center}
	\begin{figure}[t!] 
		\centering
		\includegraphics[scale=0.2]{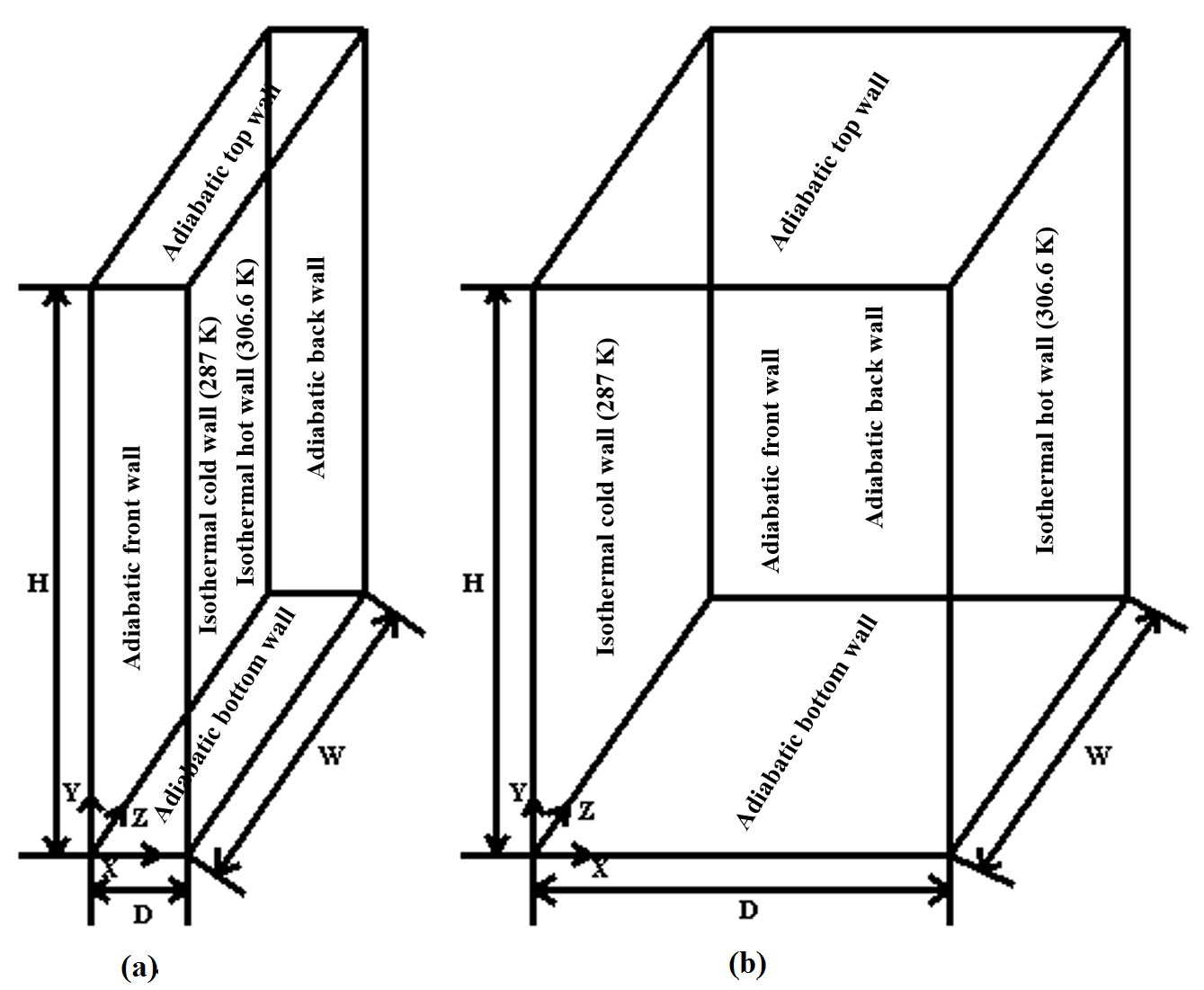}
		\caption{Computational geometries (a) Slender cavity (b) Square-base tall cavity}
		\label{fig:geo}
	\end{figure}
\end{center}

Figure \ref{fig:geo} shows the two tall cavities, i.e, slender Fig. \ref{fig:geo}(a), and square-base  Fig. \ref{fig:geo}(b) for the present study. The height (H) and the width (W) of both the cavities are same, i.e., 2.18 m and 0.52 m, respectively, whereas the depth (D) of slender and square-base cavities are 0.076 m and 0.52 m, respectively. The vertical walls separated by a distance 0.076 m and 0.52 m, respectively, in slender and square-base cavities are isothermal at temperatures $15.1^{o} C$ and $34.7^{o} C$ and the rest walls are adiabatic. The working fluid inside the cavities is the atmospheric air.  The experimental work \cite{betts2000experiments} has been performed by considering the depth of cavity as characteristic length and reported the Rayleigh number of the slender cavity $0.86\times10^{6}$ and also observed that the flow in the core of cavity was fully turbulent. However, the Rayleigh number of the order $10^6$ lies in the transition region also the direction of the gravity is parallel to the vertical walls, with these facts, the present authors believe that the height of the cavity may be considered as characteristic length. Owing to this fact, the Rayleigh number of both the cavities is $1.94\times10^{10}$. Nevertheless, to make this study consistent with the experimental work, the depth of the cavity has been considered as the characteristic length. Thus, the corresponding Rayleigh numbers of slender and square-base cavities are $0.86\times10^{6}$, and $2.75\times10^{8}$, respectively. 

Moreover, to investigate the effects of non-gray/gray radiation modeling on natural convection, the following four scenarios have been considered 

\textbf{Case A:} Pure natural convection (without radiation)

\textbf{Case B:} Combined natural convection and radiation in non-participating medium (transparent medium)

\textbf{Case C:} Combined natural convection and radiation in gray participating medium

\textbf{Case D:} Combined natural convection and radiation in non-gray medium
\section{Mathematical Formulation and Numerical Procedures}
To simulate the buoyancy driven flow with the radiation for above four scenarios, the following assumptions have been made
\begin{enumerate}
	\item The flow is steady, incompressible, and turbulent.
	\item The thermophysical properties of the fluid are constant.
	\item The medium absorbs-emits but does not scatter the radiation energy.
	\item The refractive index of the medium and walls are constant and equal to one.
	\item Boussinesq approximation is used to model the buoyancy.
	\item The medium inside cavities may behave as transparent or participating for radiative heat transfer.
	\item The participating medium may be gray or non-gray.
	\item The non-gray/gray radiative property is constant within the operating range of temperature and pressure inside the cavities.
	\item The non-gray radiative property only depends on the wavelength within the operating range of the temperature and the pressure inside the cavities.
\end{enumerate}
    with above assumptions, the continuity, the mean flow velocities and the temperature equations are following\\
    Continuity Equation\\
\begin{equation}
\label{eq:cont}
\frac{\partial \overline{u}_i}{\partial x_i}=0 
\end{equation}
Momentum Equation\\
\begin{equation}
\label{eq:momentum}
\frac{\partial \overline{u_i}~\overline{u_j}}{\partial x_j}=\frac{-1}{\rho}\frac{\partial \overline{p}}{\partial x_i}-g\beta(T-T_{ref})\delta_{i2}+\frac{1}{\rho}\frac{\partial}{\partial x_j} \Big( \mu \Big (\frac{\partial \overline{u}_i}{\partial x_j}+\frac{\partial \overline{u}_j}{\partial x_i}\Big )
-\rho\overline{u_i'u_j'}\Big)
\end{equation}
Energy Equation
\begin{equation}
\label{eq:energy}
\frac{\partial \overline{u_j}\overline{T}}{\partial x_j}=\frac{1}{\rho}\frac{\partial}{\partial x_j}\Big(\frac{k}{c_p}\frac{\partial \overline{T}}{\partial x_j}-\rho\overline{u_j' T'}\Big)-\frac{\partial q_{R,i}}{\partial x_{i}} 
\end{equation}
where $u$, $p$, $\beta$, $g$, $T$, $k$, $c_p$, $\mu$, and $\rho$ have been used in their general notations. $i$ and $j$ are the tensor indices which vary from 1 to 3 in Cartesian coordinates system  and $x$ is coordinate direction. $\delta_{i2}$ is the Kronecker delta and defined as 
\begin{equation*}
    \delta_{i2} =
    \begin{cases}
            1, &         \text{if } i=2,\\
            0, &         \text{if } i\neq 2.
    \end{cases}
\end{equation*} 

The last term of eq. (\ref{eq:momentum}) and second last term of eq. (\ref{eq:energy}) are Reynolds' stresses, and fluctuating temperature and velocity correlation terms, respectively, and modeled by SST k-$\omega$ turbulent model in the present study. The Reynolds' stresses and fluctuating velocity and temperature correlation terms are given by

\begin{equation}
-\rho \overline{{u_{i}}^{'} {u_{j}}^{'}}=\mu_{t}\Big(\frac{\partial \bar{u}_{i}}{\partial x_{j}}+\frac{\partial \bar{u}_{j}}{\partial x_{i}}\Big)-\frac{2}{3}\Big(\rho k +\mu_{t}\frac{\partial \bar{u}_{i}}{\partial x_{i}}\Big)\delta_{ij}
\end{equation}
\begin{equation}
\rho \overline{u_{i}^{'} T^{'}}=\frac{\mu_{t}}{Pr}\frac{\partial \bar{T}}{\partial x_{j}}
\end{equation}
where $\mu_{t}$ is turbulent viscosity and $'$  represents fluctuation quantities.

 The SST k-$\omega$ turbulence model is a two-equations turbulent model which involves partial-differential equations for turbulent kinetic energy $(k)$ and specific rate of dissipation $(\omega)$\cite{wilcox1998turbulence,manualANSYS}. The SST is convolution of k-$\omega$ model for the inner parts of the boundary layer, and k-$\epsilon$ for the free stream, thereby avoiding the fact that the k-$\omega$ is being sensitive to free stream. 
 
The equations for turbulent kinetic energy and specific rate of dissipation are given as
\begin{equation}
\label{eq:k}
\rho \frac{\partial}{\partial t}( k)+\rho \frac{\partial}{\partial x_i}( k \bar{u_i})=\frac{\partial}{\partial x_j}\Big(\Gamma_k\frac{\partial k}{\partial x_j}\Big)+\tilde{G_k}-Y_k
\end{equation}
\begin{equation}
\label{eq:omega}
\rho \frac{\partial}{\partial t}( \omega)+ \rho \frac{\partial}{\partial x_i}( \omega \bar{u_i})=\frac{\partial}{\partial x_j}\Big(\Gamma_\omega\frac{\partial \omega}{\partial x_j}\Big)+G_\omega-Y_\omega+D_\omega 
\end{equation}
where $\tilde{G_k}$ represents the generation of turbulence kinetic energy due to mean velocity gradients, $G_\omega$ is the generation of specific dissipation, $\Gamma_k$ and  $\Gamma_\omega$ represent the effective diffusivity of $k$ and $\omega$, respectively. $Y_k$ and $Y_\omega$ are the dissipation of $k$ and $\omega$ due to turbulence and $D_w$ represents the cross-diffusion term.

The last term of the eq. (\ref{eq:energy}) is radiation source term and calculated from the radiative intensity field and is obtained as

\begin{equation}
    \frac{\partial q_{R,i}}{\partial x_{i}} = \sum_\lambda k_\lambda(4\pi I_{b\lambda} - G_{\lambda})
\end{equation}
where $G_{\lambda}$ is the spectral irradiation and calculated from the spectral intensity field $I_\lambda$ as
\begin{equation}
    G_\lambda = \int I_\lambda d\Omega
\end{equation}

The spectral intensity $I_\lambda$ is obtained by solving the spectral radiative transfer equation (RTE) in an absorbing and emitting medium (refractive index 1) and is given as \cite{modest2013radiative,siegel2001thermal}

\begin{equation}
\label{eq:rte}
\nabla.(I_{\lambda}(\vec{r},\vec{s})\vec{s})+a_{\lambda}I_{\lambda}(\vec{r},\vec{s})=a_{\lambda}\frac{\sigma T^4}{\pi}
\end{equation}

here,  $I_{\lambda}$ is spectral intensity, $a_{\lambda}$ is spectral absorption coefficient, $\vec{r}$ and $\vec{s}$ are position and direction vectors, respectively. The above RTE is subjected to following boundary condition

\begin{equation}
I_{\lambda, w} = \epsilon_{\lambda, w} I_{b,\lambda} +\frac{1-\epsilon_{\lambda, w}}{\pi}\int_{\hat{n}.\hat{s} >0} I_{\lambda}|\hat{n}.\hat{s}|d\Omega ~~~~ for ~~~~ \hat{n}.\hat{s}<0
\end{equation}

where $\hat{n}$ and $\hat{s}$ are the unit normal area and direction vectors, respectively. $\epsilon_{\lambda,w}$ is the spectral emissivity of the wall\\

The material used in the experiments \cite{betts2000experiments} for active and passive walls are aluminimum and perspex respectively, whose gray emissivity are 0.35 and 0.9 \cite{EnggToolBox}, respectively and has been considered constant over the spectrum. Therefore, these emissivities have been used in simulations.
The conductive, radiative and total fluxes are calculated from the temperature and intensity field as,
 \begin{equation}
 q_{C}=-k\frac{\partial T}{\partial n}
 \end{equation}
 \begin{equation}
  q_{R,\lambda}=\int_{4\pi} I_{\lambda}(\hat{n}.\hat{s})d\Omega 
  \end{equation}
  
 \begin{equation}
 q_{R}=\int_{\lambda} q_{R,\lambda}d\lambda
\end{equation}  

\begin{equation}
  q_{T}=q_{C}+q_{R}
\end{equation}
 
and the corresponding, conductive, radiative, and total Nusselt number are defined as
 
\begin{equation}
Nu=Nu_{C}+Nu_{R}
\end{equation}
where
\begin{equation}
Nu_{C}=\frac{q_{C}D}{k(T_{h}-T_{c})}
\end{equation}
and
\begin{equation}
Nu_{R}=\frac{q_{R}D}{k(T_{h}-T_{c})}
\end{equation}
 $D$ is the characteristic length (depth of cavity), $k$ is thermal conductivity of the fluid. $T_h$ and $T_c$ are the hot and the cold walls temperatures, respectively.\\
 
 The three-dimensional fluid flow behavior has been visualized by Q-criterion which is the positive value of second invariant of velocity gradient tensor and is given as

\begin{equation}
    Q = \frac{1}{2}\left[\left(\mathrm{tr} \left(\nabla \textbf{u}\right)\right)^2 -\mathrm{tr}\left(\nabla\textbf{u}. \nabla\textbf{u}\right)\right]
\end{equation}

\begin{center}
	\begin{figure}[t!]
		\centering
		\includegraphics[scale=0.4]{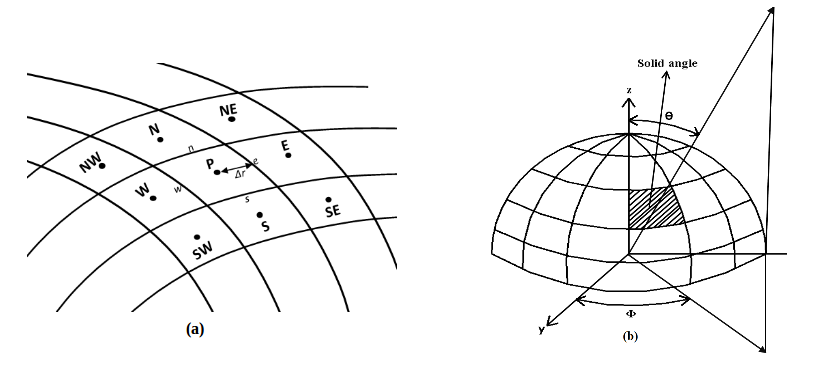}
		\caption{(a) cell arrangements for finite volume method (b) spherical discretization for RTE}
		\label{fig:cells}
	\end{figure}
\end{center}

The commercial CFD software ANSYS Fluent has been used to obtain the numerical solution of the fluid flow, the temperature, and the radiative heat transfer equations. The ANSYS Fluent uses the finite volume method (FVM) which integrates the differential equations, i.e., (\ref{eq:momentum}), (\ref{eq:energy}), (\ref{eq:k}), (\ref{eq:omega}) over a cell  and converts these equations into set of algebraic equations of the following form

\begin{equation}
    a_p\phi_p = \sum_{nb} a_{nb}\phi_{nb} + b
\end{equation}

where $\phi$, $a_p$ and $a_{nb}$ are the scalar, the central coefficients and the coefficients of the neighbouring cells, respectively, as shown in Fig \ref{fig:cells}(a) and $(b)$ is the source term. Furthermore, the radiative transfer eq. (\ref{eq:rte}) is triple integrated over wavelength, solid angle (as shown in Fig \ref{fig:cells}(b)) along with over a cell to convert into set of algebraic equations. The more details about the FVM method can be found in \cite{patankar2018numerical}. Further, the set of algebraic equations have been solved by SIMPLAC algorithm with second second order upwind scheme for momentum and energy equations and first order upwind scheme for RTE equation. The expressions of the upwind and the second order upwind schemes are given by\\

upwind scheme
\begin{equation}
    \phi_f = 
    \begin{cases}
    \phi_p & \text{if $f_\phi > 0$} \\
    \phi_{nb} & \text{if $f_\phi < 0$}
    \end{cases}
\end{equation}

second order upwind scheme
\begin{equation}
    \phi_f = 
    \begin{cases}
    \phi_p + \nabla\phi_p.\Delta r_{p,f} & \text{if $f_\phi > 0$} \\
    \phi_{nb} + \nabla\phi_{nb}.\Delta r_{nb,f} & \text{if $f_\phi < 0$}
    \end{cases}
\end{equation}
where $f_\phi$ is the either the momentum, the energy or the radiative flux at a face and $f$ indicates the face.\\
The spectral band model of the radiation module of ANSYS Fluent has been utilized for the non-gray radiative transfer simulation. The spectral band absorption coefficient of working fluid (the atmospheric air) has been calculated from HITEMP-2010 database as described  in the paragraph below.
 
The spectral band model divides the whole spectrum into certain number of bands with identification of minimum and maximum wavelength $\lambda_{min}$ and $\lambda_{max}$; respectively for each band. The intensity and the absorption coefficient are considered to be constant within a band. Thus, the eq. (\ref{eq:rte}) is modified for band model as 
\begin{equation}
\label{eq:rte:modified}
   \nabla.(I_{band}(\vec{r},\vec{s})\vec{s})+a_{band}I_{band}(\vec{r},\vec{s})= a_{band} F_{\lambda_{max}-\lambda_{min}}\frac{\sigma T^4}{\pi}
\end{equation}

where $F_{\lambda_{max}-\lambda_{min}}$ is the fractional Planck function.

\subsection{Non-Gray Gas Radiative Property Model}

The radiative property of a gas is highly dependent on wavelength, temperature, and pressure. Various databases have been developed over the last couple of  decades to record the spectroscopic parameters of a line transition for different gaseous species. Some of these databases are (1) carbon dioxide spectroscopic databank (CDSD) \cite{tashkun2011cdsd} (2) high-resolution transmission molecular absorption database (HITRAN) \cite{rothman2009hitran}  (3) high-temperature molecular spectroscopic database (HITEMP) \cite{rothman2010hitemp}. HITEMP/HITRAN database provide line transition information of a atom/molecule at 296 K and 1 atm pressure over a spectral wave-numbers range in form of 160 character length  in ASCII format. HITEMP database is the most comprehensive of all as it contains many more line transitions of a atom/molecule. An mathematical procedure needs to develop to obtain the required information from these databases at different temperature and pressure for a single specie or mixture of species. 
The atmospheric air whose molar mass composition of $N_2$, $O_2$, $CO_2$ and $H_2O$ are 75.964\%, 21\%, 0.036\% and 3\%, respectively, has been considered as the working fluid inside the natural cavity. The spectral absorption coefficient of the atmospheric air is calculated as
\begin{equation}
a_{air, \lambda}=\sum_{i = all \hspace{0.07cm} gases}^{} a_{i,\lambda}
\end{equation}

where $a_{i,\lambda}$ spectral absorption coefficient of single gas molecule and also depends on the molar mass of individual species in the mixture along with the wavelength, temperature and pressure. The nitrogen and oxygen are non-participating gases in the radiative heat transfer as these gases possess only one fundamental band and no permanent dipole moment \cite{modest2013radiative,siegel2001thermal,banwell1994fundamentals}, thus, their interaction with radiative energy is negligible, they are, therefore, considered together in calculation of spectral absorption coefficient of air, while $CO_{2}$ is taken in the same proportion as present in the atmosphere. The composition of water vapour varies from 1\% to 4\% in the atmosphere \cite{bartwal18china}, whereby only 3\% of water vapour has been considered in the present calculations. The current authors have performed these calculations to obtain spectral absorption coefficients in an isothermal and homogeneous medium for single species $CO_{2}$ \cite{bartwal17bits}, $H_{2}O$ \cite{bartwal17iitm}, and the atmospheric air \cite{bartwal18china}. 

\begin{table}[t!]
	\centering
	\addtolength{\tabcolsep}{-6pt}
	\caption{Band width, fractional Planck function and absorption coefficient of the atmospheric air in 25 bands for entire spectrum}
	\label{band25}
	\begin{tabular}{|c|c|c|c|c|}
		\hline
		Band No. & $\lambda_{min}$($\mu m$) & $\lambda_{max}$($\mu m$) & $F_{\lambda_{max}-\lambda_{min}}$(298 K) & Absorption coefficient ($m^{-1}$) \\ \hline
		1           & 1.07                     & 2.52                     & 0.000006                                 & 0.012                            \\ \hline
		2           & 2.52                     & 2.59                     & 0.000003                                 & 0.920                            \\ \hline
		3           & 2.59                     & 2.73                     & 0.000012                                 & 1.738                            \\ \hline
		4           & 2.73                     & 2.82                     & 0.000013                                 & 0.855                            \\ \hline
		5           & 2.82                     & 4.17                     & 0.003062                                 & 0.002                            \\ \hline
		6           & 4.17                     & 4.21                     & 0.000269                                 & 0.318                            \\ \hline
		7           & 4.21                     & 4.31                     & 0.000747                                 & 1.348                            \\ \hline
		8           & 4.31                     & 4.37                     & 0.000503                                 & 0.122                            \\ \hline
		9           & 4.37                     & 5.30                     & 0.014296                                 & 0.047                            \\ \hline
		10          & 5.30                     & 5.63                     & 0.008510                                 & 0.421                            \\ \hline
		11          & 5.63                     & 6.96                     & 0.053487                                 & 1.945                            \\ \hline
		12          & 6.96                     & 22.47                    & 0.711011                                 & 0.058                            \\ \hline
		13          & 22.47                    & 25.00                    & 0.041972                                 & 0.491                            \\ \hline
		14          & 25.00                    & 37.73                    & 0.102087                                 & 4.094                            \\ \hline
		15          & 37.73                    & 39.21                    & 0.005776                                 & 0.145                            \\ \hline
		16          & 39.21                    & 41.66                    & 0.008134                                 & 21.219                           \\ \hline
		17          & 41.66                    & 43.47                    & 0.005064                                 & 1.535                            \\ \hline
		18          & 43.47                    & 46.51                    & 0.007075                                 & 15.184                           \\ \hline
		19          & 46.51                    & 47.61                    & 0.002193                                 & 8.067                            \\ \hline
		20          & 47.61                    & 50.00                    & 0.004212                                 & 38.774                           \\ \hline
		21          & 50.00                    & 55.55                    & 0.007507                                 & 6.073                            \\ \hline
		22          & 55.55                    & 58.82                    & 0.003323                                 & 24.296                           \\ \hline
		23          & 58.82                    & 60.60                    & 0.001550                                 & 5.187                            \\ \hline
		24          & 60.60                    & 117.64                   & 0.015706                                 & 16.645                           \\ \hline
		25          & 117.64                   & 181.81                   & 0.002122                                 & 7.187                            \\ \hline
	\end{tabular}
\end{table}

The operating temperature range inside the cavities is in the range 15.1-34.7 $^o$C and flow is buoyancy driven, thus, the pressure variation is almost negligible, therefore the variation of absorption coefficient is considered with wavelength only in the present study. With this fact, the spectral absorption coefficient is being calculated at average temperature of 298 K and 1 atm over entire spectrum of thermal radiation. First, the entire spectrum is divided into 25 bands and the Planck mean average absorption coefficient of each bands is calculated as shown in Table \ref{band25}. Along with the absorption coefficient, fractional Planck function for the corresponding band is also depicted in table.  The absorption coefficient of these bands have been used in the band radiation model of the ANSYS Fluent. In order to minimize computational resource required for radiation model, the simulation is also performed for 13 and 9 bands of entire spectrum and corresponding absorption coefficients and fractional Planck functions are depicted in Table \ref{band13} and \ref{band9}, respectively. The gray absorption coefficient is obtained by calculating the Planck mean average coefficient over entire spectrum and that is turned out to be 1.471 $m^{-1}$.     

\begin{table}[t!]
	\centering
	\addtolength{\tabcolsep}{-6pt}
	\caption{Band width, fractional Planck function, and absorption coefficient of the atmospheric air in 13 bands for entire spectrum}
	\label{band13}
	\vspace{0.3cm}
	\begin{tabular}{|c|c|c|c|c|}
		\hline
		Band No. & $\lambda_{min}$($\mu m$) & $\lambda_{max}$($\mu m$) & $F_{\lambda_{max}-\lambda_{min}}$(300 K) & Absorption coefficient ($m^{-1}$) \\ \hline
		1           & 1.07                     & 2.52                     & 0.000006                                 & 0.012                            \\ \hline
		2           & 2.52                     & 2.82                     & 0.000030                                 & 1.266                            \\ \hline
		3           & 2.82                     & 4.17                     & 0.003062                                 & 0.002                            \\ \hline
		4           & 4.17                     & 4.37                     & 0.001521                                 & 0.765                            \\ \hline
		5           & 4.37                     & 5.30                     & 0.014296                                 & 0.047                            \\ \hline
		6           & 5.30                     & 6.96                     & 0.061997                                 & 1.737                            \\ \hline
		7           & 6.96                     & 22.47                    & 0.711011                                 & 0.058                            \\ \hline
		8           & 22.47                    & 28.16                    & 0.080406                                 & 0.877                            \\ \hline
		9           & 28.16                    & 37.73                    & 0.063652                                 & 5.788                            \\ \hline
		10          & 37.73                    & 46.51                    & 0.026052                                 & 11.067                           \\ \hline
		11          & 46.51                    & 50.00                    & 0.006405                                 & 28.164                           \\ \hline
		12          & 50.00                    & 54.05                    & 0.005739                                 & 5.961                            \\ \hline
		13          & 54.05                    & 181.81                   & 0.024471                                 & 15.399                           \\ \hline
	\end{tabular}
\end{table}

\begin{table}
	\centering
	\addtolength{\tabcolsep}{-6pt}
	\caption{Band width, fractional Planck function, and absorption coefficient of the atmospheric air in 9 bands for entire spectrum}
	\label{band9}
	\vspace{0.3cm}
	\begin{tabular}{|c|c|c|c|c|}
		\hline
		Band No. & $\lambda_{min}$($\mu m$) & $\lambda_{max}$($\mu m$) & $F_{\lambda_{max}-\lambda_{min}}$(300 K) & Absorption coefficient ($m^{-1}$) \\ \hline
		1           & 1.07                     & 2.52                     & 0.000006                                 & 0.012                            \\ \hline
		2           & 2.52                     & 2.82                     & 0.000030                                 & 1.266                            \\ \hline
		3           & 2.82                     & 4.17                     & 0.003062                                 & 0.002                            \\ \hline
		4           & 4.17                     & 4.37                     & 0.001521                                 & 0.765                            \\ \hline
		5           & 4.37                     & 5.30                     & 0.014296                                 & 0.047                            \\ \hline
		6           & 5.30                     & 6.96                     & 0.061997                                 & 1.737                            \\ \hline
		7           & 6.96                     & 22.47                    & 0.711011                                 & 0.058                            \\ \hline
		8           & 22.47                    & 37.73                    & 0.144059                                 & 4.272                            \\ \hline
		9           & 37.73                    & 181.81                   & 0.062668                                 & 16.143                           \\ \hline
	\end{tabular}
\end{table}

\section{Sensitivity Analysis of Selections}
The numerical solution involving non-gray radiation model is sensitive to number of grids, directions as well as number of spectral bands used in the simulation. It is therefore, the tests of sensitivity analysis for the spatial and directional grids and number of spectral bands on the final results have been performed. The $SST$ $k-\omega$ turbulent model which is a low Reynolds number turbulent model, has been used to simulate the turbulent flow. The value of $y^+$ for the first grid point has been kept $\sim1$ for both the cavities.

\subsection{Grids' Sensitivity Test}

The spatial and angular grids sensitivity analysis tests have been performed for both the cavities, i.e., the slender and the square base cavities.

\subsubsection{Slender Cavity}

\begin{figure}[t!]
	\begin{center}
		\includegraphics[scale=0.2]{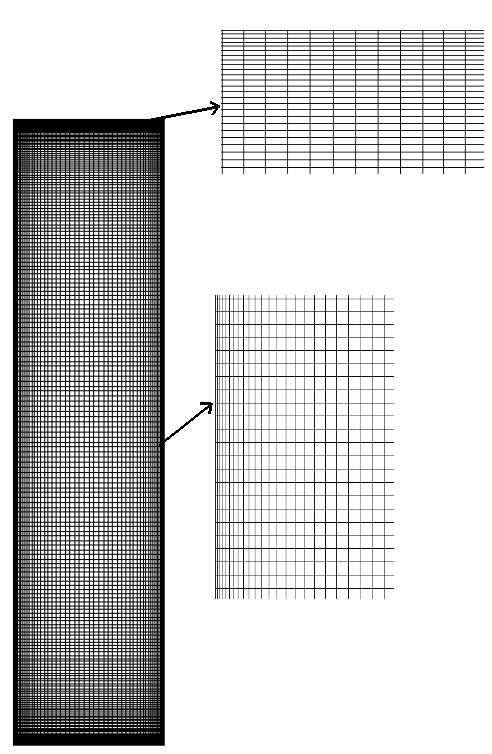}
		\caption{Non-uniform grid for numerical simulation in cavities}
		\label{fig:grid}
	\end{center}
\end{figure}

\begin{figure*}[b!]
$\begin{array}{cc}
    \includegraphics[width=0.4\textwidth]{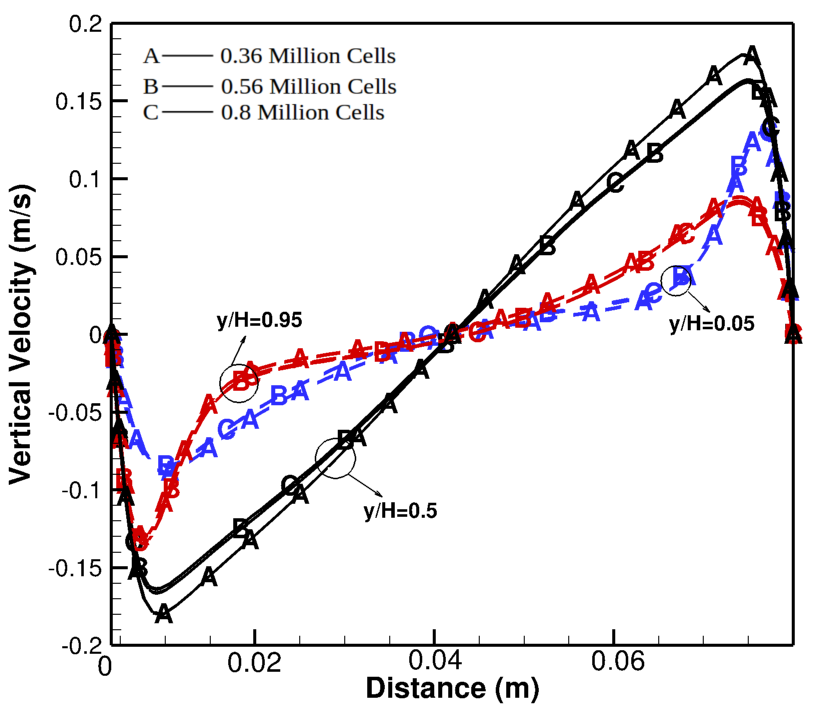} &
    \includegraphics[width=0.4\textwidth]{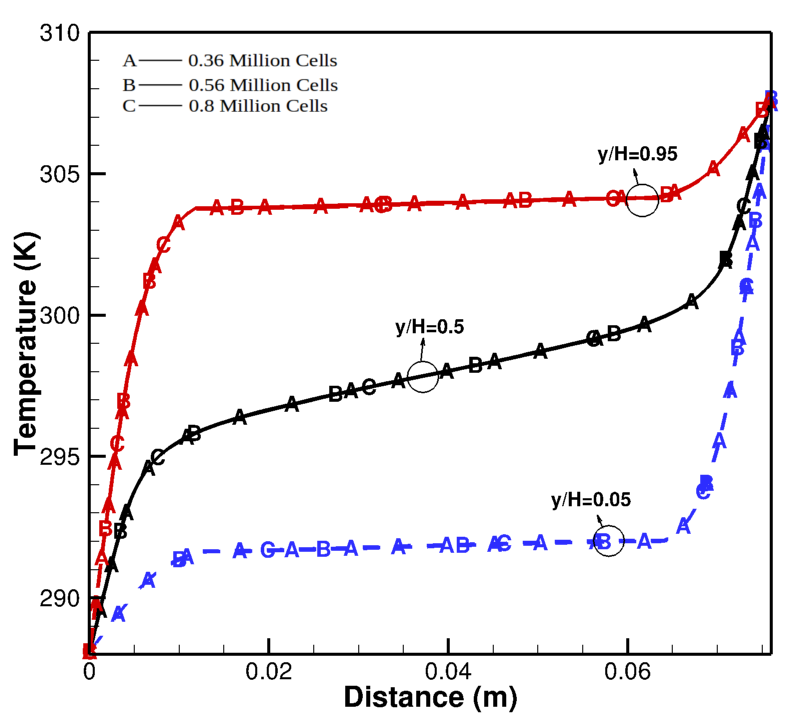}\\
    \mathrm{(a)} & \mathrm{(b)} \\
\end{array}$
\caption{\label{fig:indGrid} Spatial grids' sensitivity test study on the vertical velocity and the temperature profile for simulation of pure convection in slender cavity}

\end{figure*}

The spatial grids' sensitivity test is carried out for the optimal number of cells for simulation of natural convection in the slender cavity. A three-dimensional geometry similar to experimental setup of Betts and Bukhari \cite{betts2000experiments} has been constructed, and the simulations are performed for 0.36, 0.56 and 0.81 million cell numbers for the pure convection problem. A non-uniform grids with finer grids near the boundaries and little coarse grids in the interior with stretching ratio 0.85, has been generated as shown in Fig. \ref{fig:grid}.

Figures \ref{fig:indGrid} (a) and (b) show the vertical velocity and the temperature variations, respectively along the line on the mid plane perpendicular to the active walls at various non-dimensional heights y/H = 0.05, 0.5 and 0.95 for different number of cells. A major difference appears in the vertical velocity curve between the simulated solution with 0.36 and 0.56 million cells, however, differences is negligible between 0.56 and 0.8 million cells. The temperature profiles, (figure \ref{fig:indGrid}(b)) at various non-dimensional heights of the cavity do not show any sensitivity with the number of cells. Thus, the case with 0.56 million cells has been chosen for the further study.

\begin{table}[!b]
\centering
\caption{Total Heat flux ($W/m^{2}$) report of slender tall cavity for gray medium for different angular discretizations}
\label{angular_ind_Slender}
\begin{tabular}{|c|c|c|c|}
\hline
\multirow{2}{*}{Walls} & \multicolumn{3}{c|}{ $n_{\theta} \times n_{\phi}$ } \\ \cline{2-4} 
 & 2$\times$2 & 3$\times$3 & 4$\times$4 \\ \hline
Isothermal Cold & -58.88 & -58.30 & -58.33 \\ \hline
Isothermal Hot & 58.88 & 58.30 & 58.33 \\ \hline
\end{tabular}
\end{table}

The effect of angular discretization on total heat flux on the isothermal the active walls for slender cavity with 0.56 million spatial discretization is shown in Table \ref{angular_ind_Slender}. The absolute percentage difference between the first and the second angular discretization is 0.9\% , whereas in second and third angular discretization is 0.05\%. Thus, finally $n_{\theta} \times n_{\phi} = 3 \times 3$ angular space is selected for the study of the present problem.

\subsubsection{Square-base Tall Cavity}

\begin{figure*}[t!]
$\begin{array}{cc}
    \includegraphics[width=0.4\textwidth]{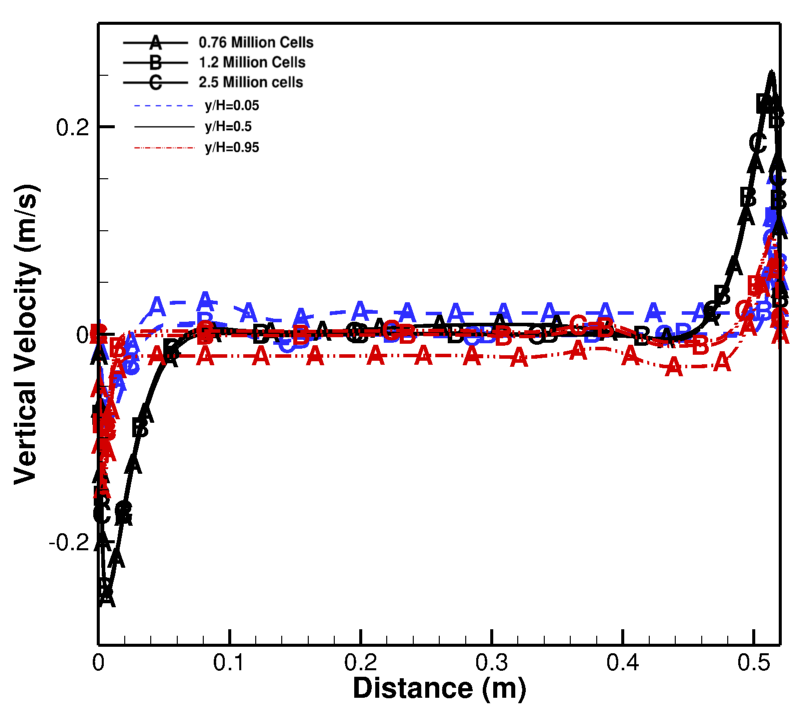} &
    \includegraphics[width=0.4\textwidth]{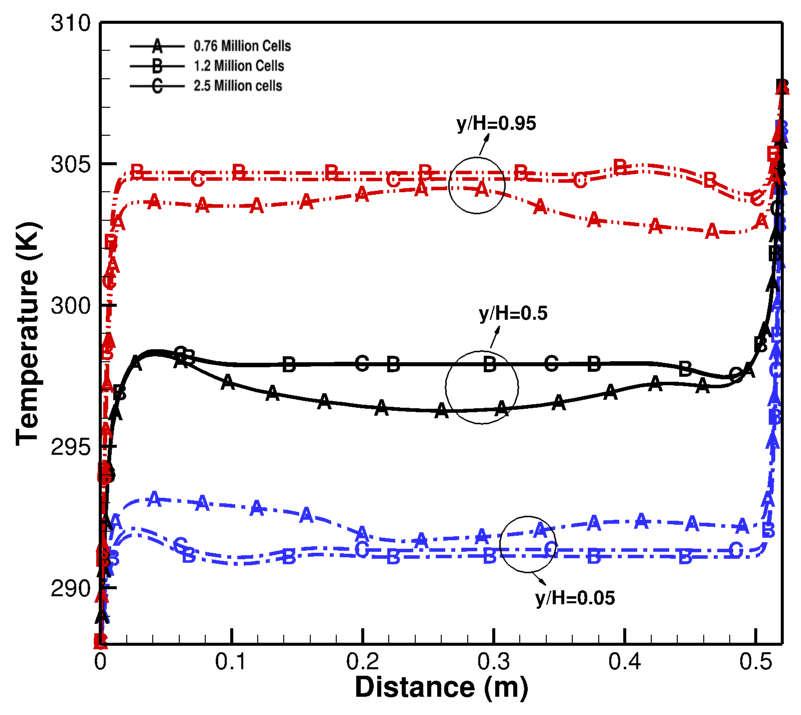}\\
    \mathrm{(a)} & \mathrm{(b)} \\
\end{array}$
\caption{\label{fig:indSqu}Spatial grids' sensitivity test study on the vertical velocity and the temperature profile for simulation of pure convection in square base cavity}
\end{figure*}

Figures \ref{fig:indSqu}(a) and (b) depict the variation of the vertical velocity and the temperature, respectively, along the line on a plane perpendicular to active walls at non-dimensional heights y/H = 0.05, 0.5 and 0.95 in pure convection simulation with various grids of 0.76, 1.2 and 2.5 million cells for the square-base tall cavity. The 0.76 and 1.2 million cells show marked difference on the results of the vertical velocity and the temperature, however, 1.8 and 2.5 million cells simulation do not show any difference on the results of vertical velocity and temperature. Therefore, a case with 1.2 million cell has been selected for the simulation of natural convection in the square base cavity.

\begin{table}[!t]
\centering
\caption{Total heat flux ($W/m^{2}$) report of square-base tall cavity for gray medium for different angular discretizations}
\label{angular_ind_square}
\begin{tabular}{|c|c|c|c|}
\hline
\multirow{2}{*}{Walls} & \multicolumn{3}{c|}{$n_{\theta} \times n_{\phi}$} \\ \cline{2-4} 
 & 2 $\times$ 2 & 3 $\times$ 3 & 4 $\times$ 4 \\ \hline
Isothermal Cold & -50.1 & -50.16 & -50.08 \\ \hline
Isothermal Hot & 50.1 & 50.16 & 50.08 \\ \hline
\end{tabular}
\end{table}

The effect of angular discretization on total heat flux on the active walls of the square-base tall cavity  with 1.2 million spatial discretization is shown in Table \ref{angular_ind_square}. The absolute percentage difference between the first and second angular discretization is 0.11\% , whereas in second and third angular discretization is 0.15\%.  The difference in fluxes for chosen directions is quite small and also considering the study of sensitivity in direction selection in slender cavity, finally $n_{\theta} \times n_{\phi} = 3 \times 3$ angular space is selected for the study of present problem.

\subsection{Sensitivity test for selection of number of bands}
The whole thermal spectrum (1.07$\mu$m - 180$\mu$m) of thermal radiation has been divided into 9, 13 and 25 bands and the corresponding fractional Planck function and Planck average absorption coefficient of the respective bands have also been calculated as shown in Tables \ref{band25}, \ref{band13} and \ref{band9}, respectively. Inorder to enhance the effect of radiation mode of heat transfer, so that, the sensitivity in the bands selection may be more visible, the simulation of natural convection with band radiation model for  slender cavity has been performed with wall emissivity 1. The total flux (conduction + radiation) on the walls of slender cavity (for three categories of bands selection, i.e, 9, 13 and 25 has been reported in table \ref{heatFluxSlender}). The difference of flux between 9 and 13 bands simulation is approximately 0.7 $W/m^2$ and between 13 and 25 band is approximately 0.1 $W/m^2$. Thus, 13 bands category has been selected for the study of non-gray radiation interaction with natural convection in the tall cavities.

\begin{table}[t!]
	\centering
	\addtolength{\tabcolsep}{-5pt}
	\caption{ Total heat flux report of the slender tall cavity for different number of bands for radiation simulation with emissivity 1 for all walls}
	\label{heatFluxSlender}
	\begin{tabular}{|c|c|c|c|}
		\hline
		\multirow{2}{*}{Walls} & \multicolumn{3}{c|}{Total Heat Flux ($W/m^{2}$)} \\ \cline{2-4} 
		& 9 Bands         & 13 Bands        & 25 Bands       \\ \hline
		Adiabatic back wall                   & 0               & 0               & 0              \\ \hline
		Adiabatic bottom wall                 & 0               & 0               & 0              \\ \hline
		Adiabatic front wall                  & 0               & 0               & 0              \\ \hline
		Isothermal left wall                   & -137.183        & -137.857       & -137.958       \\ \hline
		Isothermal right wall                  &  -137.183        & 137.857       & 137.958     \\ \hline
		Adiabatic top wall                    & 0               & 0               & 0             \\ \hline
	\end{tabular}
\end{table}

\section{Results and Discussion}

The geometry of two cavities as shown in Fig.1, have been used to investigate the effects of non-gray/gray radiative heat transfer on natural convection at low operating temperature range. First, the simulation results for four scenarios of radiative modeling for the slender cavity have been compared with the experimental results, whereupon, a comprehensive analysis of fluid flow and heat transfer have been presented. Further, to augment the effect of radiation, the distance between the active walls which is 0.076 m for slender cavity case, has been increased to 0.52 m (Fig.1(b)) such that base of the cavity becomes square, and named this cavity as square-base cavity. The results of both the cavities are presented in following sections:

\subsection{Slender Cavity}
Due to the small distance (D = 0.076 m) between the active walls compared to other dimensions (W = 0.52 m, H = 2.18 m), it has been reported that the fluid flow and heat transfer phenomena are two-dimensional at the middle plane perpendicular to the active walls, however, phenomena could be complex beyond this plane and radiation may play a major role. It is therefore, a three-dimensional simulation is performed on the actual experimental geometry (slender cavity) and the fluid flow and the heat transfer characteristics are investigated in the following sections for this cavity.

\subsubsection{Fluid Flow Characteristics}
\begin{figure}[b!]
	\begin{center}
		\includegraphics[width =10cm,height=6cm]{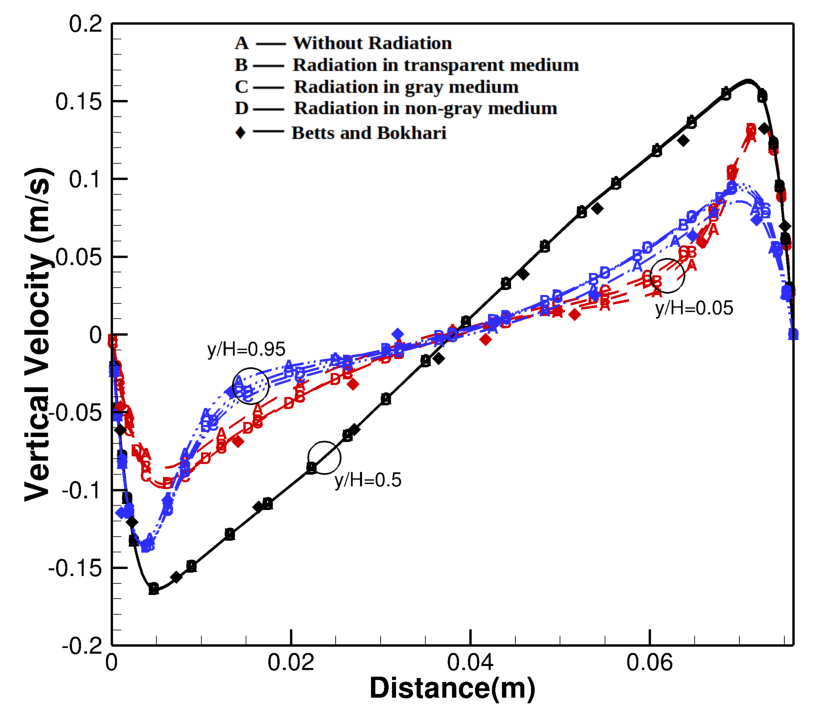}
		\caption{Variation of vertical velocity along the line on the mid plane of the cavity perpendicular to the active walls at different heights}
		\label{fig:verVelSl}
	\end{center}
\end{figure}

The variations of vertical velocity (Y-velocity) along a line on the middle plane perpendicular to active walls at the non-dimensional heights of y/H = 0.05, 0.5, and 0.95 are shown in Fig. \ref{fig:verVelSl}.  This velocity graph also includes the vertical velocity variation for four scenarios of the radiation modeling as mentioned in section 2 along with the experimental results. The difference in vertical velocities is not found at the middle of the cavity for four scenarios of radiation modeling, and also matches well with the experimental results, however little difference is observed at other locations of the cavity away from the walls, i.e., at the top and the bottom location of the cavity. This difference is negligible near to the walls. The vertical velocity changes its sign from negative to positive and vice-versa at the middle of the cavity (asymmetry) and increases in the respective directions away from the mid point. This is direct consequence of the onset of vertically upward and downward flows due to buoyancy that have been setup near to the hot and the cold wall, respectively. The maximum vertical velocity is 0.16 m/s which is achieved close to both the active walls at y/H = 0.5, and decreases to 0.04 m/s, and 0.09 m/s at the top (y/H = 0.95), and bottom (y/H = 0.05) of the cavity, respectively, at the cold wall and this is achieved at a distance 0.005m from the wall. The vertical velocity in pure convection case appears little less than to vertical velocities in other three cases beyond at a distance of 0.1 m and upto 0.03 m from the cold and the hot wall at y/H=0.95 and 0.05, respectively. This difference in vertical velocity between pure convection and other cases is around 10$\%$. There is no distinct boundary layer seen at middle height, however, the thickness of boundary layer is 0.03 m at the top and at the bottom of the cavity. The maximum vertical velocity values on the cold wall at y/H=0.05 for pure convection, radiation in participating, gray and non-gray medium are 0.075, 0.08, 0.082 and 0.083 m/s, respectively while these values on hot wall are equal i.e 0.9 m/s for radiation cases at mid height of the cavity , this value is 0.17m/s for the pure convection case and these trend are same on the hot wall as well.

\begin{figure}[t!]
	\begin{center}
		\centering
		\includegraphics[width =11cm,height=7cm]{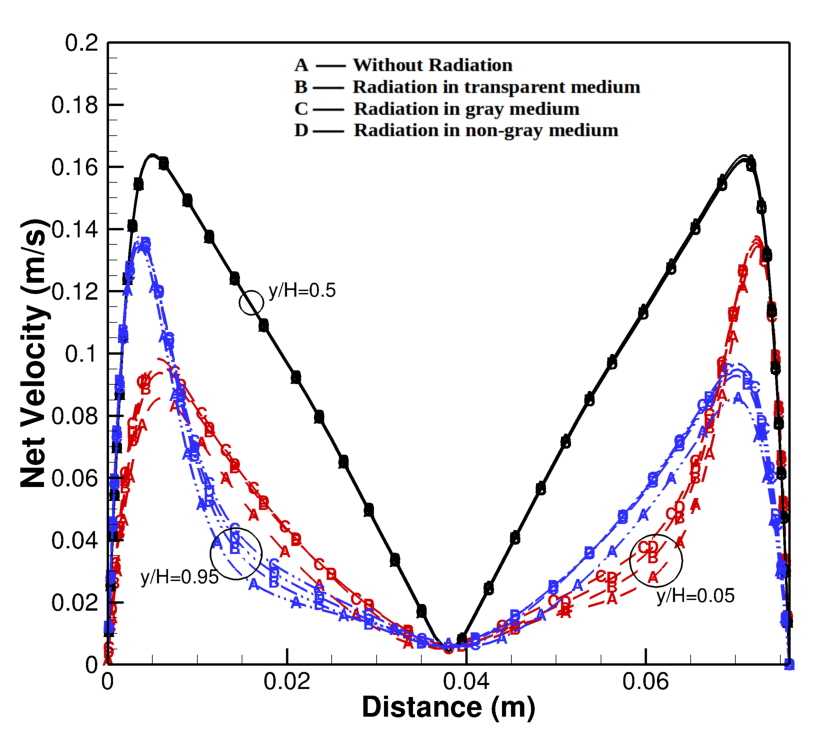}
		\caption{Variation of net velocity at various heights on mid-plane perpendicular to active walls of the cavity}
		\label{fig:netVelSl}
	\end{center}
\end{figure}

\begin{figure*}[]
$\begin{array}{cc}
    \includegraphics[width=0.5\textwidth]{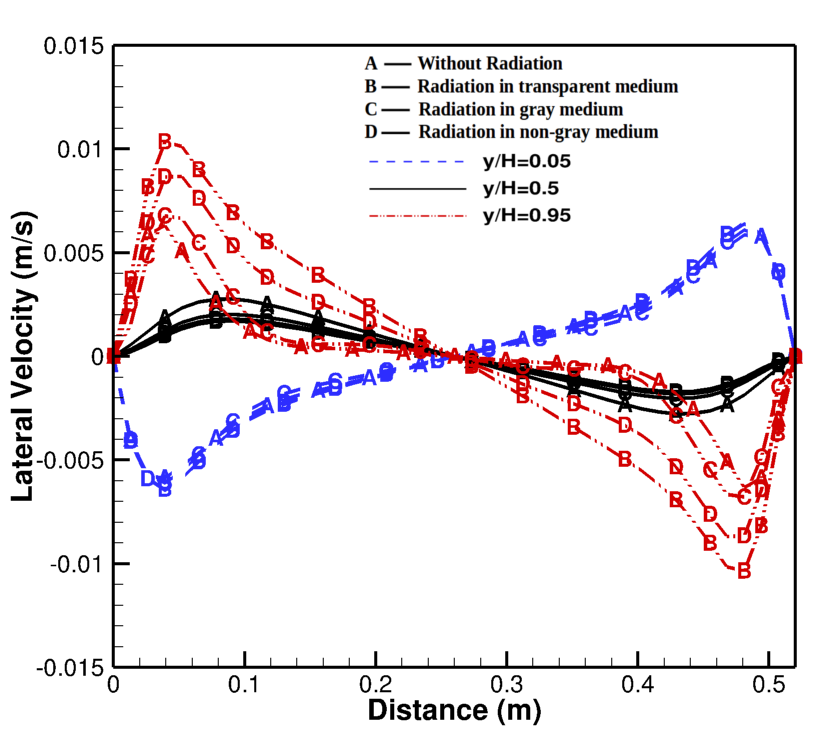} &
    \includegraphics[width=0.5\textwidth]{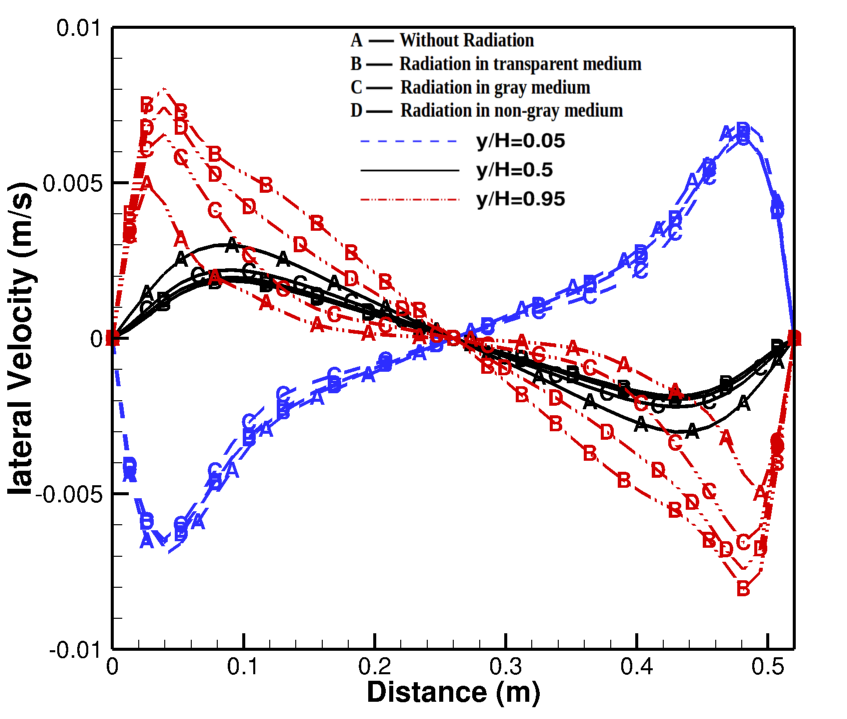}\\
    \mathrm{(a)~~ (near~~ to~~ the~~ cold~~ wall)} & \mathrm{(b)} \\
    \multicolumn{2}{c}{\includegraphics[width=0.5\textwidth]{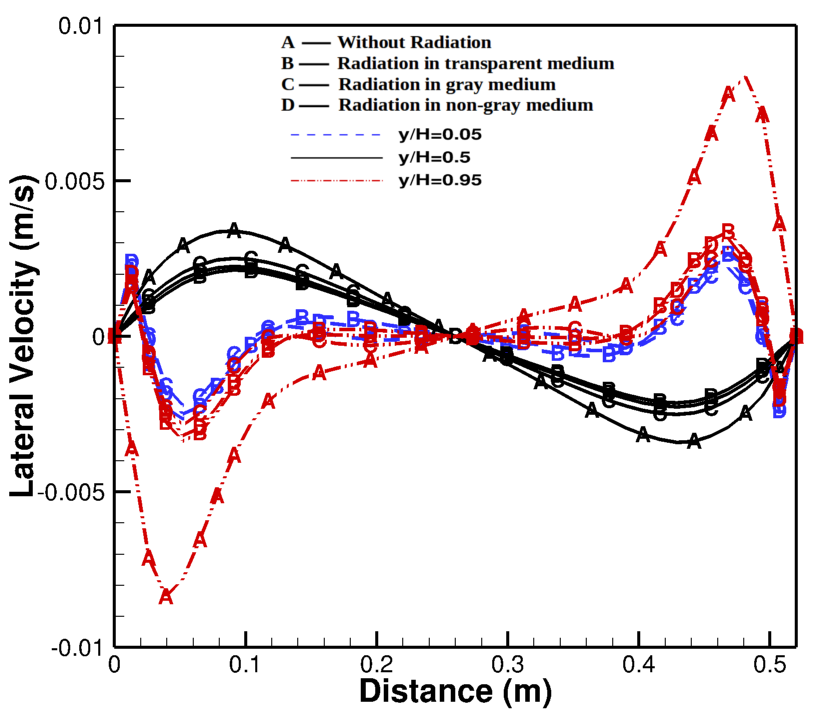}}\\
    \multicolumn{2}{c}{(c)}\\
    \includegraphics[width=0.5\textwidth]{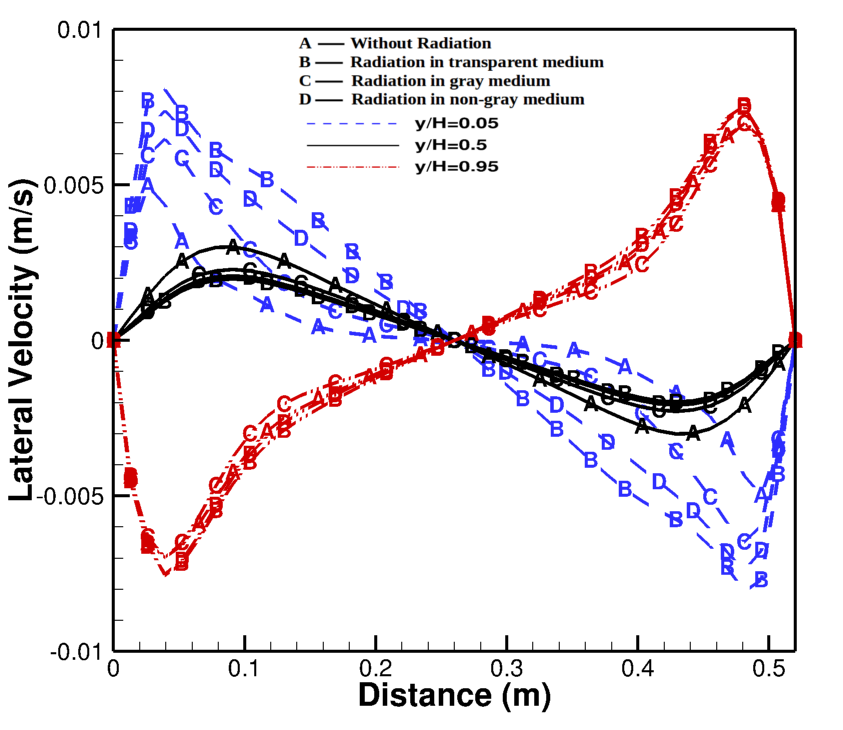} &
    \includegraphics[width=0.5\textwidth]{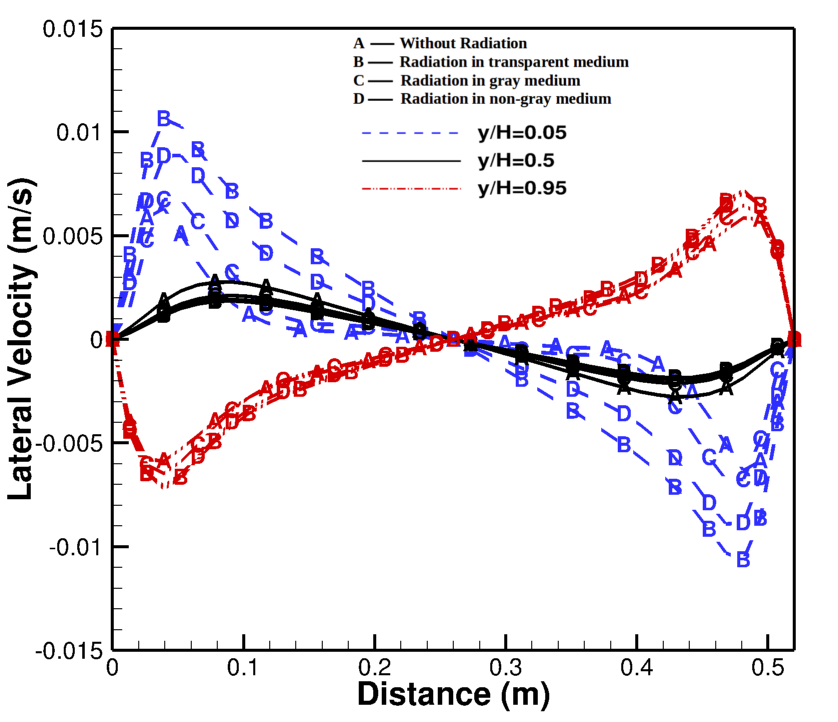}\\
    (d) & (e)\\
\end{array}$
\caption{\label{fig:sleVz}Variation of the lateral velocity ($V_{z}$) along the z-direction at various heights on the planes parallel to the isothermal walls at a distance (a) 0.01 m (b) 0.015 m (c) 0.038 m (d) 0.061 m (e) 0.066 m from the cold wall}
\end{figure*}

The net velocities, i.e., $u = \sqrt{{u_{1}}^2+{u_{2}}^2+{u_{3}}^2}$ along the line on the mid plane perpendicular to the active walls at three vertical distances, i.e., y/H = 0.05, y/H = 0.5, and y/H = 0.95  are depicted in Fig. \ref{fig:netVelSl} for four different scenarios of radiation modeling. The net velocity in all four scenarios is almost identical at the mid height of the cavity, i.e., y/H = 0.5, however, some differences have been found away from walls at the top and the bottom of the cavity. The fluid is not at rest at middle of the cavity but has same minimum net velocity over entire vertical height for all scenarios and this minimum velocity is 0.005m/s. The maximum velocity at the mid height of the cavity is equal and attained at a distance of 0.008m from both the walls. The values of maximum net velocity near to hot wall at height of y/H=0.05, 0.5 and 0.95 are 0.1, 0.17 and 0.14 m/s, respectively which reveal that the fluid starts rising from the bottom of hot wall and accelerate upto mid height of the cavity and decelerate afterwords. The reverse of this is visible near to the cold wall. It appears that nature of fluid flow near to the hot wall is mirror image of fluid flow near to the cold wall. This is owing to fact that maximum net velocity near to hot wall at y/H=0.05 is same as maximum net velocity at y/H=0.95 near to the cold wall and maximum net velocities are is equal at y/H=0.5 near to the hot and the cold walls. The boundary layer thickness at the lower and upper part of the cavity is same and equal to 0.0032 m which grows to half of the cavity depth, i.e, 0.038 m at mid height of the cavity. The four scenarios of radiation modeling only show difference at the lower part of the cavity near to the hot wall and lower part of cavity near to the cold wall. The pure convection shows the minimum net velocity among all scenarios over depth of the cavity.

\begin{figure*}[]
$\begin{array}{cc}
    \includegraphics[width=5cm,height=9cm]{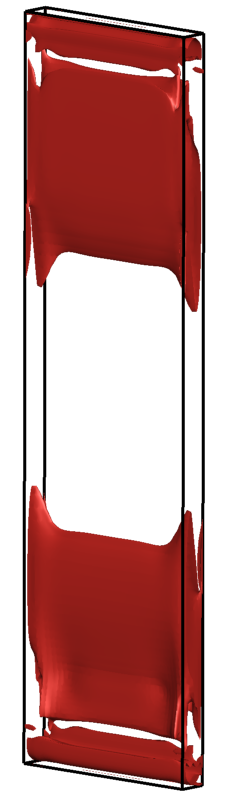} &
    \includegraphics[width=5cm,height=9cm]{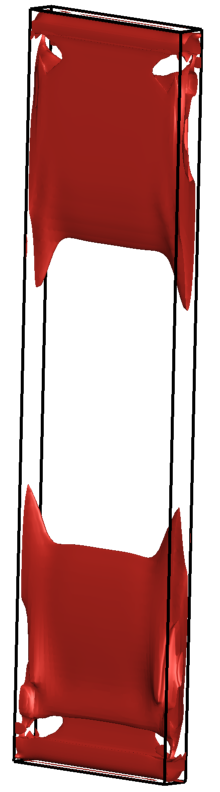}\\
    \mathrm{(A)} & \mathrm{(B)} \\
    \includegraphics[width=5cm,height=9cm]{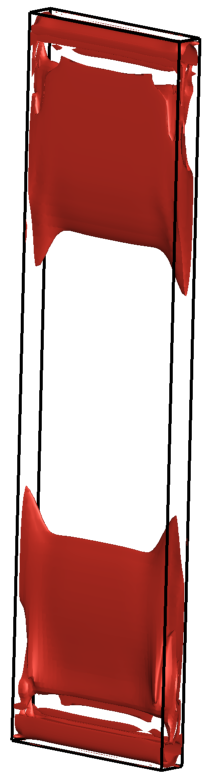} &
    \includegraphics[width=5cm,height=9cm]{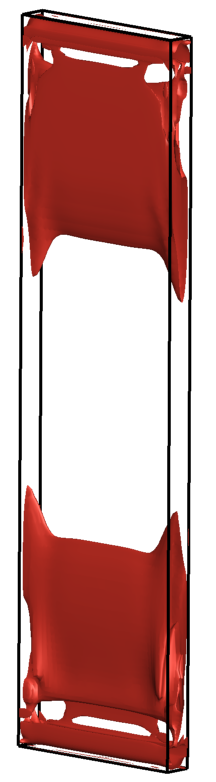}\\
    (C) & (D)\\
\end{array}$
\caption{\label{fig:qSlen}Isosurface of Q criterion for a value of 0.03 for pure convection (A) pure convection, and (B) combined convection and radiation in transparent medium, (C) gray medium, (D) non-gray medium}
\end{figure*}

The depth of cavity is quite small compared to other dimensions, it was therefore, reported that two-dimensional flow would prevail at the mid plane (z = 0.26 m) of the cavity \cite{betts2000experiments}, whereas flow may be three-dimensional in other parts of the cavity. Thus, it will be informative to know the variation of lateral velocity ($u_{3}$) in the third direction. Figure \ref{fig:sleVz} shows the variation of lateral velocity ($u_{3}$) in the z-direction at different heights of the cavity, i.e., y/H = 0.05, y/H = 0.5, and y/H = 0.95 on a plane parallel to active walls at distances (a) x = 0.01 m (b) x = 0.015 m (c) x = 0.038 m (d) x = 0.061 m (e) x = 0.066 m from the cold wall for the four scenarios of radiation modeling in natural convection. These distances have been chosen due to fact that the vertical velocity ($u_{y}$) (Fig. \ref{fig:verVelSl})  shows maximum variation within 0.015 m distance from the active walls. Although, the lateral velocity ($u_{3}$) is one order less than the vertical and net velocities (Fig. \ref{fig:verVelSl} and \ref{fig:netVelSl}, respectively), however, it has quite significant variation over the width of cavity (w), specially, at the top and the bottom of the cavity, i.e., y/H = 0.05, and y/H = 0.95. The lateral velocity is anti-symmetric about the center point of the graph in all the cases, nevertheless, it is quite prominent near to the active walls (Fig \ref{fig:sleVz}(a) and \ref{fig:sleVz}(e)), and  its magnitude decreases towards the center of the cavity. 
The lateral velocity magnitude is comparatively high at top (y/H = 0.95) of the cavity near to the cold wall, (see Fig.\ref{fig:sleVz}(a)) while the similar behaviour is observed at y/H = 0.05 near to the hot wall as shown in Fig. \ref{fig:sleVz}(e). The change in the radiative behaviour of the medium significantly affects the lateral velocity. The maximum lateral velocity is found in case of radiation in transparent medium, i.e., absorption coefficient is zero, whereas, minimum is for pure convection case at the top and the bottom of the cavity. Nevertheless, the lateral velocity is significantly prominent at the top and the bottom of the cavity, while it is significantly low or negligible at mid-height of the cavity, also at mid depth of the cavity.

Figure \ref{fig:qSlen} depicts the iso-surface of Q criterion for a value of 0.03. The Q criterion is positive second invariant of strain rate tensor. It can be further conceptualized as the vorticity magnitude exceeds the magnitude of strain rate, this means that it represents the rotation of the fluid. It is interesting to note that exactly similar but opposite iso-surface of Q criterion exists at the top of the cavity near to the hot wall and at the bottom of the cavity near to the cold wall. There is no apparent difference in the iso-surface of the Q criterion for the four considered scenarios of radiation modeling, i.e., pure convection, combined convection and radiation in transparent, gray, and non-gray medium. This may be due to small distance between active walls that constrict the fluid flow. The vertices are not present anywhere else in the cavity except near to lower and upper corners of the cold and the hot walls, respectively, thus the flow is shear dominant  in the slender cavity.

\subsubsection{Heat Transfer Characteristics}

\begin{figure}[t!]
	\begin{center}
		\includegraphics[width =9cm,height=6cm]{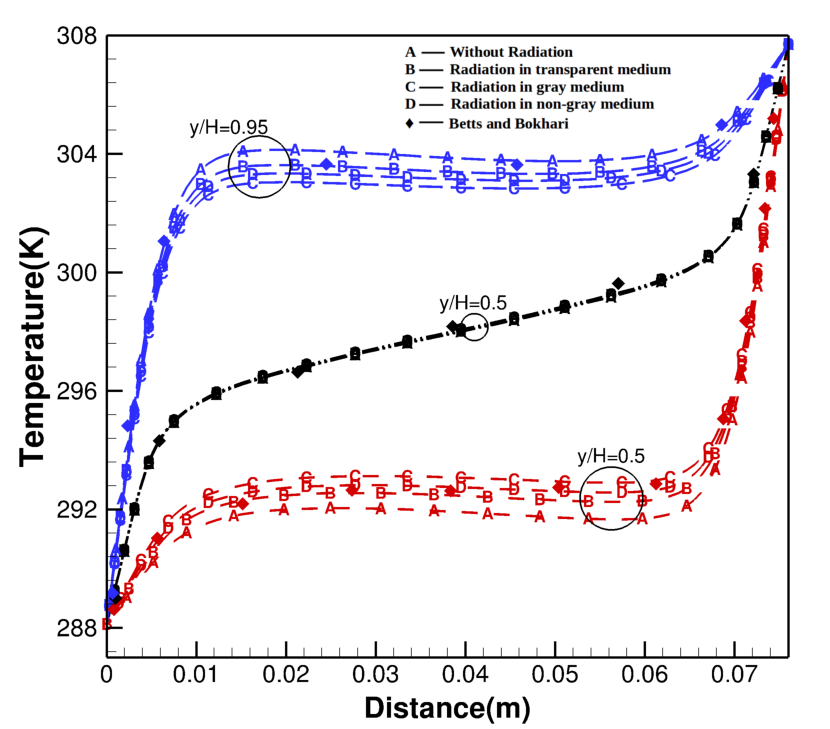}
		\caption{Variation of temperature along a line on mid plane of the cavity perpendicular to active walls at different heights}
		\label{fig:tempV}
	\end{center}
\end{figure}

The temperature variations along a line on the middle plane (z = 0.26 m) perpendicular to the active walls at different heights of the slender cavity are illustrated in Fig. \ref{fig:tempV} for four scenarios of radiation modeling along with the experimental results. The temperature profile is same along the line at mid height of the cavity, however, difference in temperature profile appears in the four scenarios at the top and the bottom of the cavity. The highest and the lowest temperature are found in the pure convection case at the top and the bottom of the cavity, respectively. The reverse trend is observed in the case of gray medium. The non-gray medium for the radiation predicts the temperature profile between those two extreme cases of the pure convection and the radiation in gray medium. The experimental temperature profile also closely matches with the non-participating medium case, however, temperature profile for non-gray medium case is very close to the transparent medium temperature profile. The temperature is the highest at the top and the lowest at the bottom, this is result of rise and descend of the hot and the cold fluid due to buoyancy. There is a almost 2 K temperature difference at the top and the bottom of the cavity between the pure convection and radiation in the gray medium cases. 

\newpage

\begin{figure*}[]
$\begin{array}{cc}
    \includegraphics[width=5cm,height=9cm]{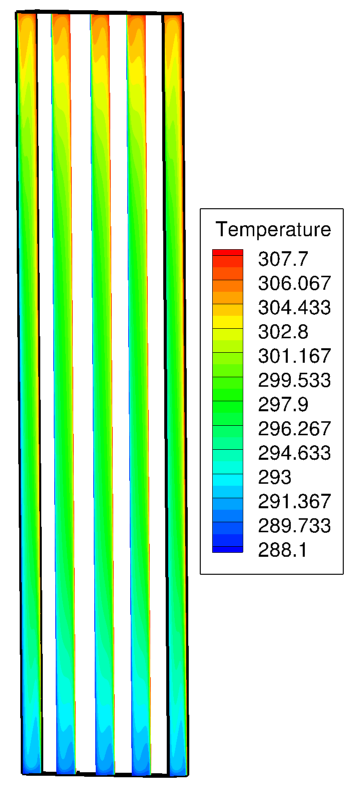} &
    \includegraphics[width=5cm,height=9cm]{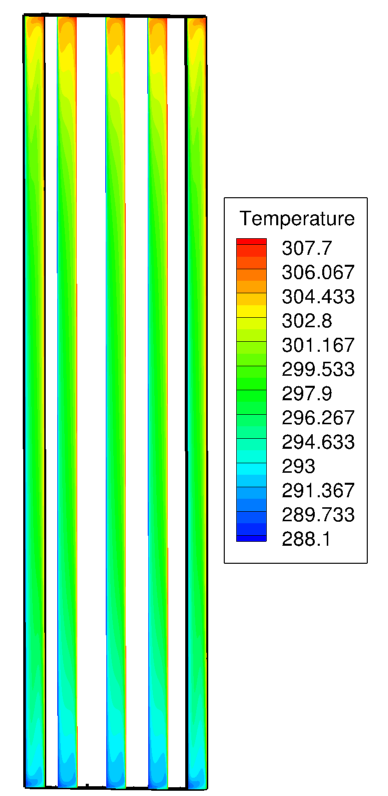}\\
    \mathrm{(A)} & \mathrm{(B)} \\
    \includegraphics[width=5cm,height=9cm]{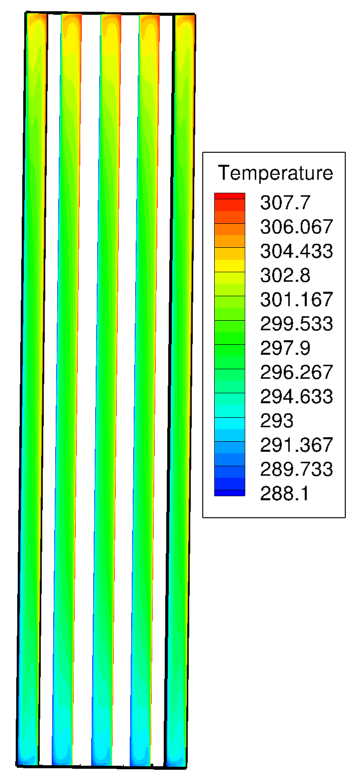} &
    \includegraphics[width=5cm,height=9cm]{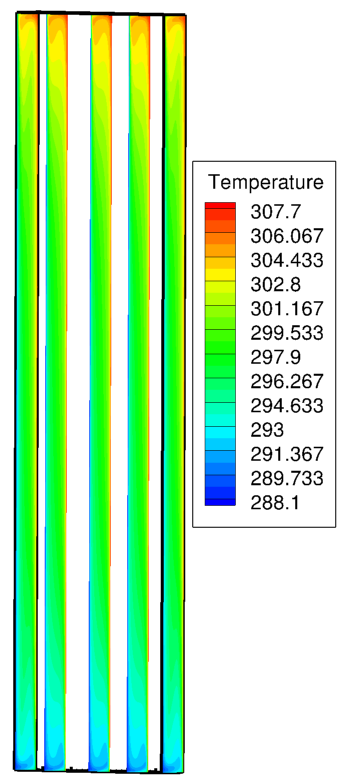}\\
    (C) & (D)\\
\end{array}$
\caption{\label{fig:tempSlen}Temperature (K) contours at various planes perpendicular walls for (A) pure convection, and (B) combined convection and radiation in transparent medium, (C) gray medium, (D) non-gray medium}
\end{figure*}

Figure \ref{fig:tempSlen} shows the temperature contours at the various planes perpendicular to the active walls for the four scenarios of radiation modeling. As we have seen in previous section, fluid flow was only getting affected at the top and the bottom of the cavity, whereas, the core of the cavity remains unaffected in four scenarios, these trends are also expected in the temperature contours since the fluid flow affects the temperature field. Figures \ref{fig:tempSlen}(a-d) indicate the presence of the relatively hotter and colder zones at the top, and the bottom of the cavity, respectively. Furthermore, the space occupied by these hotter and colder zones are changing with the behaviour of the medium for radiation, i.e, no radiation to radiation in non-gray medium. On comparing  Fig. \ref{fig:tempSlen}(a) and (b), the region over which the hot and the cold zones at the top and the bottom of the cavity are spread, has decreased and that decreases further in case of radiation in gray medium Fig. \ref{fig:tempSlen}(c), whereas, a little increase in the spread of hot region is observed in case of the radiation in non-gray medium Fig. \ref{fig:tempSlen}(d).

\begin{figure*}[]
$\begin{array}{cc}
    \includegraphics[width=5cm,height=9cm]{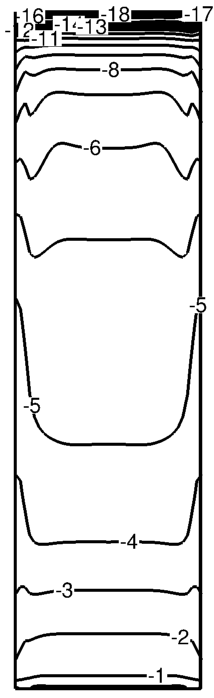} &
    \includegraphics[width=5cm,height=9cm]{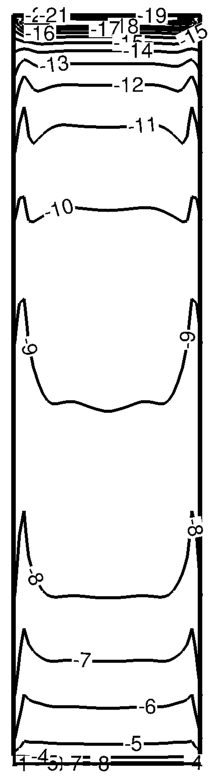}\\
    \mathrm{(A)} & \mathrm{(B)} \\
    \includegraphics[width=5cm,height=9cm]{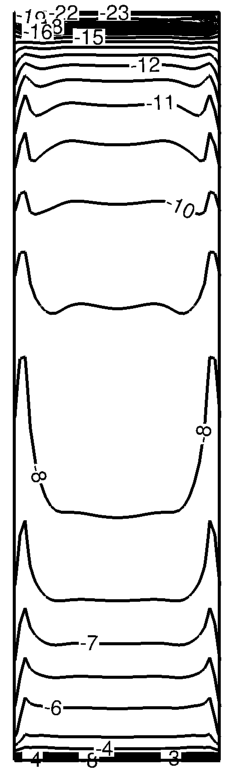} &
    \includegraphics[width=5cm,height=9cm]{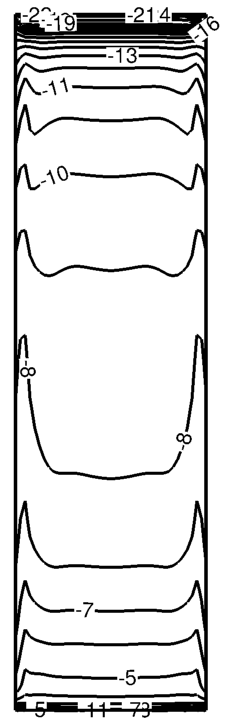}\\
    (C) & (D)\\
\end{array}$
\caption{\label{fig:nussSlenCold}Total Nusselt number contours on the cold wall (287 K) for (A) pure convection, and (B) combined convection and radiation in transparent medium, (C) gray medium, (D) non-gray medium }
\end{figure*}

\begin{figure*}[]
$\begin{array}{cc}
    \includegraphics[width=5cm,height=9cm]{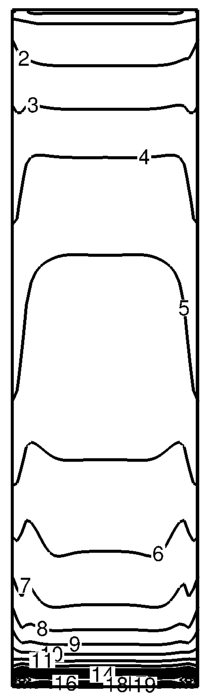} &
    \includegraphics[width=5cm,height=9cm]{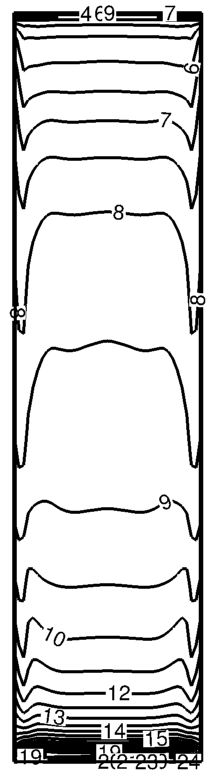}\\
    \mathrm{(A)} & \mathrm{(B)} \\
    \includegraphics[width=5cm,height=9cm]{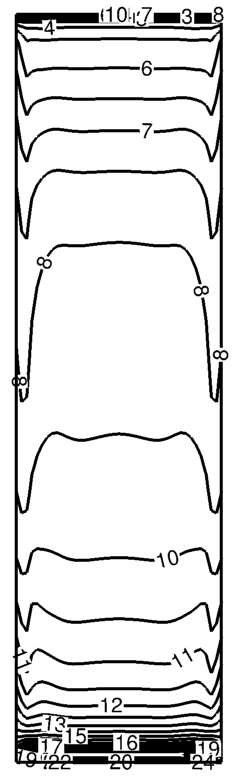} &
    \includegraphics[width=5cm,height=9cm]{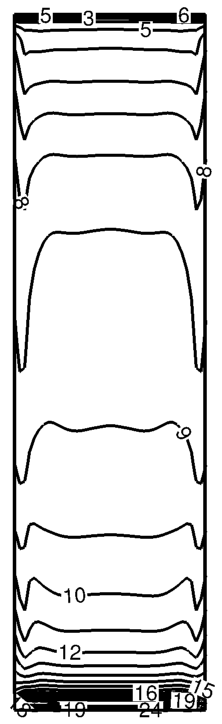}\\
    (C) & (D)\\
\end{array}$
\caption{\label{fig:nussSlenHot}Total Nusselt number contours on the hot wall (306.6 K) for(A) pure convection, and (B) combined convection and radiation in transparent medium, (C) gray medium, (D) non-gray medium}
\end{figure*}

\begin{table}[t!]
	\begin{center}
	\label{avgTempSlen}
		\caption{Average temperature inside the slender tall cavity for four scenarios of radiation modeling}
	\begin{tabular}{|c|c|c|}
		\hline
		S.No. & Case     & Average temperature \\ \hline
		1    & Case (A) & 297.90 K        \\ \hline
		2    & Case (B) & 297.93 K        \\ \hline
		3    & Case (C) & 297.93 K        \\ \hline
		4    & Case (D) & 297.93 K        \\ \hline
	\end{tabular}
\end{center}
\end{table} 

The absorption of radiation energy may lead to increase in the average temperature inside the cavity, however, it seems that there is no apparent change in the average temperature inside the slender cavity for present study as shown in Table 5. There is only 0.03 K increase in average temperature for all radiation cases than the pure convection case. This small change in the average temperature may be owing to the fact that very little absorption of radiation energy happen because of mainly two reasons, (i)  small distance between the active walls (ii) fractional Planck function is small in those bands where absorption coefficients are high and vice-versa (Please refer to Table \ref{band13}).

The total Nusselt number contours at the cold wall for the various cases considered for the study have been shown in Fig. \ref{fig:nussSlenCold}. The Nusselt number vary significantly from the top to the bottom on the cold wall and the maximum and the minimum Nusselt number are found on top and bottom of the wall, respectively. These values of Nusselt number are 20.9 and 0.2, respectively, for pure convection case, 23.9 and 2.9 for transparent medium case; 25.2 and 2.9 for gray medium case and 24.5 and 2.9 for non-gray medium case. The amount of increase in the maximum and the minimum values of the Nusselt numbers with the inclusion of radiation is almost remain same. However, there is small increase in the maximum Nusselt number for gray medium case compared to transparent medium case. While it decreases little for non-gray case in comparison with gray medium case. The similar, but opposite, i.e, now maximum and minimum Nusselt numbers are found on the bottom and the top of the hot wall, respectively. The contours of the Nusselt number for four scenarios of the radiation modeling shown in Fig.\ref{fig:nussSlenHot}.

Table \ref{heatFluxSlenR} represents the area average conductive and radiative heat fluxes ($W/m^{2}$) on both the active and the passive walls of the slender cavity for four scenarios of radiation modeling. The comparison is also made with the experimental work of another paper by Dafa Alla and Betts\cite{DafaExperimental}. The conductive and the radiative heat fluxes obtained from the present numerical simulations closely matches to the experimental results, with approximate 8\% error. This could be attributed to many uncertainties in both the measurements and the numerical methods. The conductive heat flux (CHF) remains same for all the cases, moreover, the radiative heat flux (RHF) is almost same order of the conductive heat flux for cases involving radiation. Also, the radiative flux also remains same for all cases involving the radiation model. This might be due to fact that absorption of radiation energy happened in very small amount because of small distance between the active walls. The total heat flux is zero on the adiabatic walls (top, bottom, front, and back), thus the radiative heat flux is equal and opposite to the conductive heat flux. The total wall heat flux which is sum of the conductive and the radiative heat fluxes, is same on both the active walls for all cases. One thing to notice is that the conduction/radiation flux is much smaller on the vertical adiabatic walls compared to horizontal adiabatic walls. Further, this flux is same on both the vertical adiabatic wall while this flux is higher on  the bottom adiabatic wall compared to the top adiabatic wall. Though, conduction/radiation flux remains same on the active walls for all scenarios of radiation modeling, but this is not true on adiabatic walls. The conduction/radiation flux is the highest for transparent medium case and the lowest for radiation in gray medium on the passive walls, while this value of flux falls between above two cases for radiation in non-gray medium. This may be owing to fact that emissivity ($\epsilon_w = 0.35$) of active walls are much lower than the adiabatic walls ($\epsilon_w = 0.9$). 

\begin{table}
\centering
\addtolength{\tabcolsep}{-7pt}
\caption{Heat flux ($W/m^{2}$) report of slender tall cavity for (a) pure convection, and (B) combined convection and radiation in transparent medium, (C) gray medium, (D) non-gray medium}
\label{heatFluxSlenR}
\begin{tabular}{|c|c|c|c|c|c|c|c|c|c|}
\hline
\multirow{2}{*}{Walls} & Case(A) & \multicolumn{2}{c|}{Case(B)} & \multicolumn{2}{c|}{Case(C)} & \multicolumn{2}{c|}{Case(D)} & \multicolumn{2}{c|}{\begin{tabular}[c]{@{}c@{}}Dafa Alla and Betts \cite{DafaExperimental}\end{tabular}} \\ \cline{2-10} 
 & CHF & CHF & RHF & CHF & RHF & CHF & RHF & CHF & RHF \\ \hline
Isothermal Cold & -33.722 & -33.978 & -24.268 & -34.032 & -24.272 & -34.129 & -24.128 & -34.5 & -24.5 \\ \hline
Isothermal Hot & 33.722 & 33.978 & 24.268 & 34.131 & 24.173 & 33.941 & 24.316 & 34.5 & 24.5 \\ \hline
Adiabatic Top & 0 & -13.704 & 13.703 & -11.105 & 11.105 & -12.463 & 12.463 & 0 & 0 \\ \hline
Adiabatic Bottom & 0 & 15.07 & -15.07 & 12.146 & -12.146 & 13.697 & -13.697 & 0 & 0 \\ \hline
Adiabatic Front & 0 & 0.789 & -0.789 & 0.539 & -0.539 & 0.684 & -0.684 & 0 & 0 \\ \hline
Adiabatic Back & 0 & 0.789 & -0.789 & 0.539 & -0.539 & 0.684 & -0.684 & 0 & 0 \\ \hline
\end{tabular}
\end{table}

\subsection{Square-Base Tall Cavity}

As appeared in section 4.1, that the radiation has considerable effect on the heat transfer characteristics in the slender cavity even though the average operating temperature range is quite small, nevertheless, this effect was little on the fluid flow, this could be due to small distance between the active walls, i.e., 0.076 m. The radiation is a long distance phenomenon and may have considerable effect on the fluid flow if the active walls are at sufficient distance apart. It may also alter the fluid flow characteristics inside the cavity, hence, the distance between the active walls, i.e., width of the cavity has been increased and make equal to the depth of the slender cavity. Now, the base of the tall cavity become square ($0.52 m \times 0.52 m$), hence, named as square-base tall cavity and rest conditions of the cavity are kept same as of the slender cavity. The fluid flow and the heat transfer characteristics in this square base tall cavity have been analyzed in following sections for all scenarios of radiation modeling.

\subsubsection{Fluid Flow Characteristics}

\begin{figure*}[t!]
$\begin{array}{cc}
    \includegraphics[width=6cm,height=4cm]{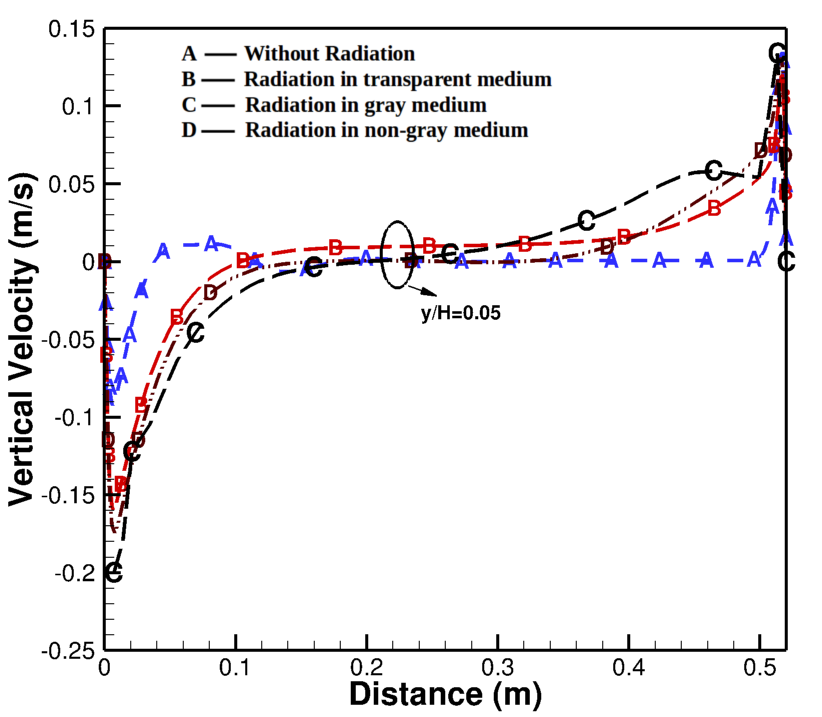} &
    \includegraphics[width=6cm,height=4cm]{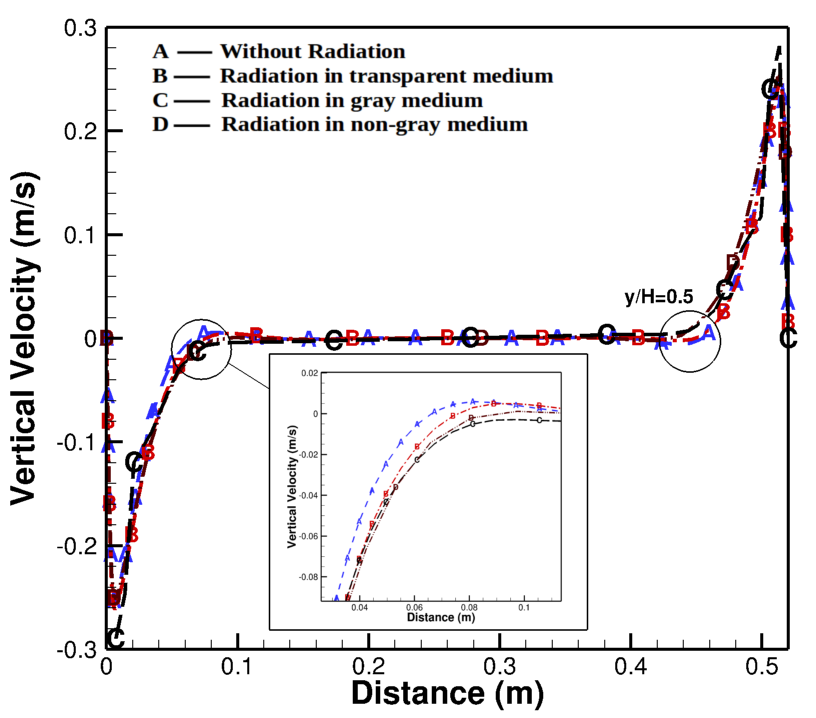}\\
    \mathrm{(A)} & \mathrm{(B)} \\
   \multicolumn{2}{c}{\includegraphics[width=6cm,height=4cm]{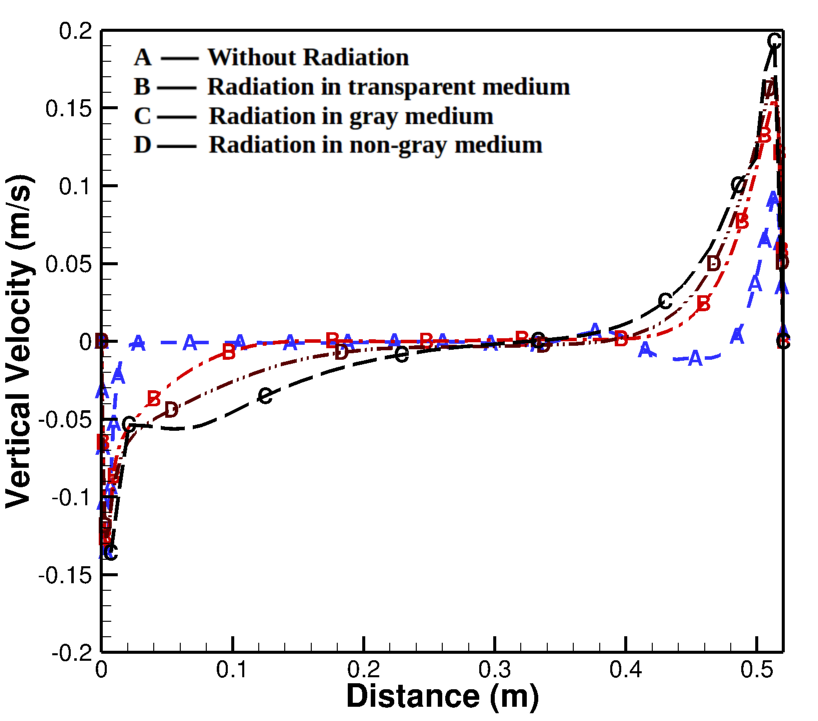}}\\
    \multicolumn{2}{c}{(c)}\\
\end{array}$
\caption{Variation of the vertical velocity ($V_{y}$) along the lines at the heights (A) y/H=0.05, (B) y/H=0.5 and (C) y/H=0.95 on the mid plane perpendicular to isothermal walls}
 \label{Fig:verVSq}
\end{figure*}

\begin{figure*}[t!]
$\begin{array}{cc}
    \includegraphics[width=6cm,height=4cm]{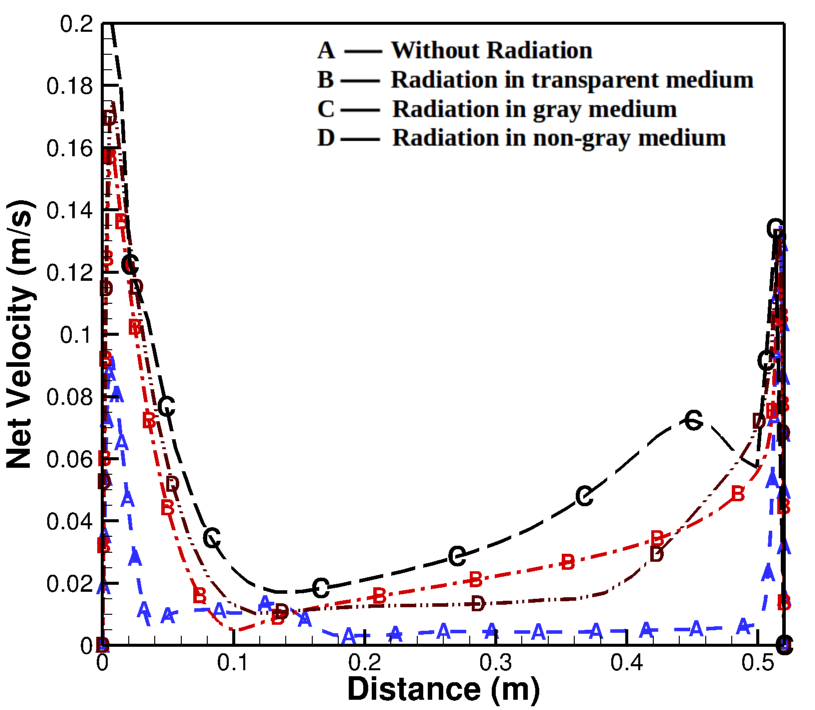} &
    \includegraphics[width=6cm,height=4cm]{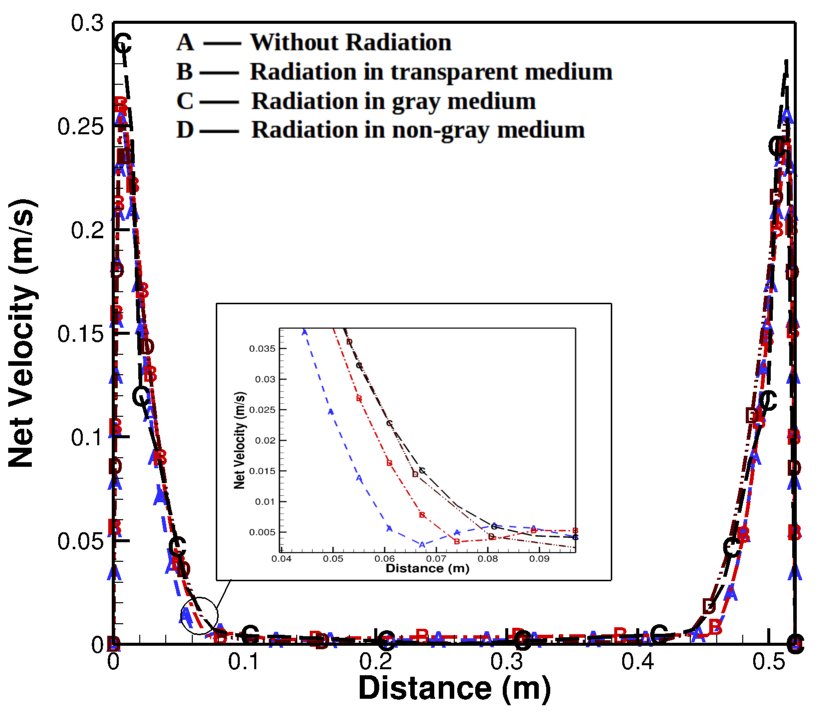}\\
    \mathrm{(A)} & \mathrm{(B)} \\
   \multicolumn{2}{c}{\includegraphics[width=6cm,height=4cm]{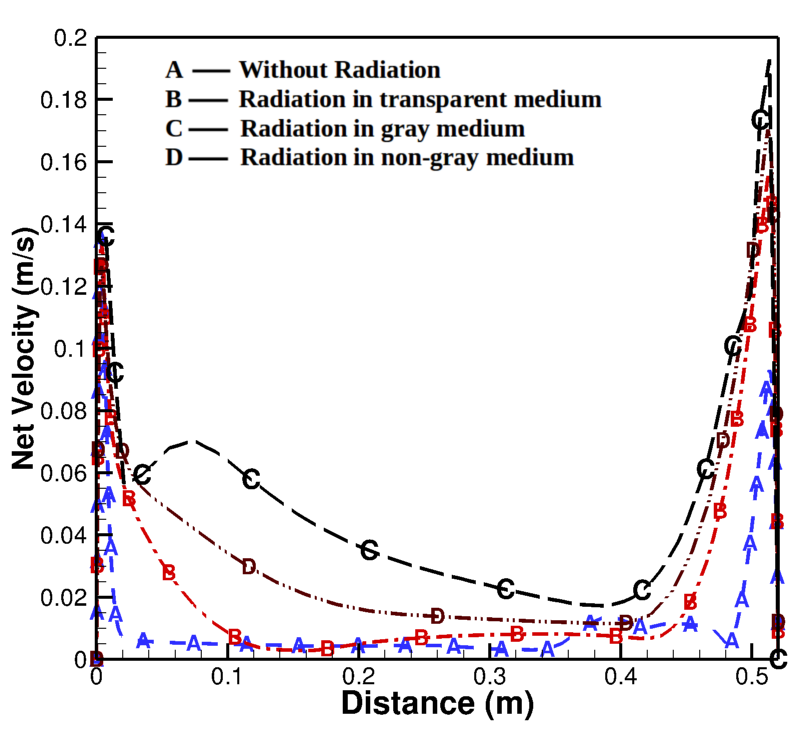}}\\
    \multicolumn{2}{c}{(c)}\\
\end{array}$
\caption{Variation of the net velocity along the line at the heights (A) y/H=0.05 (B) y/H=0.5 and (C) y/H=0.95 on the mid-plane perpendicular to isothermal walls of the cavity}
\label{Fig:netVelSq}
\end{figure*}

The variation of vertical velocity ($u_{2}$) along the line at various non-dimensional heights (A) y/H=0.05, (B) y/H=0.5 and (C) y/H=0.95 on the mid plane perpendicular to the active walls of the cavity is shown in Fig. \ref{Fig:verVSq}. The vertical velocity is only significant near to the hot and the cold walls, while it is almost zero in the core of the cavity. As fluid rises from the hot wall and descends from the cold wall, the boundary layer grows from the bottom of the cavity at the hot wall, and from the top of the cavity at the cold wall. The velocity profiles within the boundary layers are different for case of without and with radiation models, at the top and the bottom of the cavity, while, this is same at the mid height of the cavity. The thickness of boundary layer at y/H=0.05 in pure convection, radiation in transparent, gray and non gray medium cases on the cold wall are 0.1, 0.1 , 0.15 and 0.12 m, respectively. Whereas these thicknesses on the hot walls are 0.02, 0.1, 0.25 and 0.2 m, respectively. The reverse trend on boundary layer thicknesses near to the top (y/H=0.95) of the cavity appears as seen in fig. (\ref{Fig:netVelSq}C). Furthermore, these boundary layers become same for all cases and also on the cold and the hot walls of cavity at mid height of the cavity. The thicknesses of boundary layers is 0.08 m for all scenarios. Moreover, the vertical velocity is found to be the highest in the case of radiation in gray medium, followed by non-gray medium, transparent medium and the least in pure convection case. One interesting fact to notice is that barring the top and the bottom of the cavity, the fluid is at rest in the most part of the cavity.

Figure \ref{Fig:netVelSq} illustrates the net velocity, i.e, $u = \sqrt{{u_{1}}^{2}+{u_{2}}^{2}+{u_{3}}^{2}}$ along the line at various heights (A) y/H=0.05, (B) y/H=0.5 and (C) y/H=0.95 on the mid plane perpendicular to the active walls of the cavity. The radiation influences the net velocity as well as the hydrodynamic boundary layer thickness based on the net velocity, especially only at the top and the bottom of the cavity, while its influence is negligible at the mid height of the cavity. There is no distinct boundary layer seen at the top and the bottom of the cavity for all cases of radiation, however, the fluid is almost at rest away from the walls for the pure convection case. It appears that there is appreciable fluid motion at the top and bottom of the cavity. A distinct boundary layer appears at mid height of the cavity. The thickness of this boundary layer is 0.08 m and is almost same for all scenarios of radiation modeling and also on both the active walls. The maximum net velocity is higher at the bottom on the cavity near the cold wall (see Fig. \ref{Fig:netVelSq}A) while it is higher at the top near to the hot wall at the top of the cavity (see Fig. \ref{Fig:netVelSq}C). The maximum net velocity is found to be 0.26 $ms^{-1}$ near to both the active walls for gray medium case at the mid height of the cavity. One interesting thing to notice that there is no fluid motion in the core of the cavity.

\begin{figure*}[]
$\begin{array}{cc}
    \includegraphics[width=0.5\textwidth]{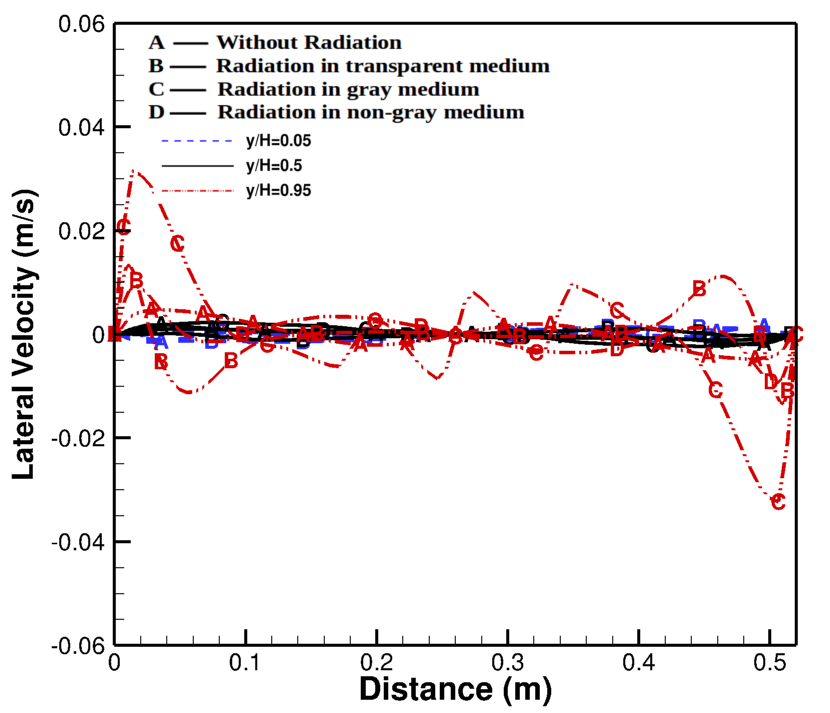} &
    \includegraphics[width=0.5\textwidth]{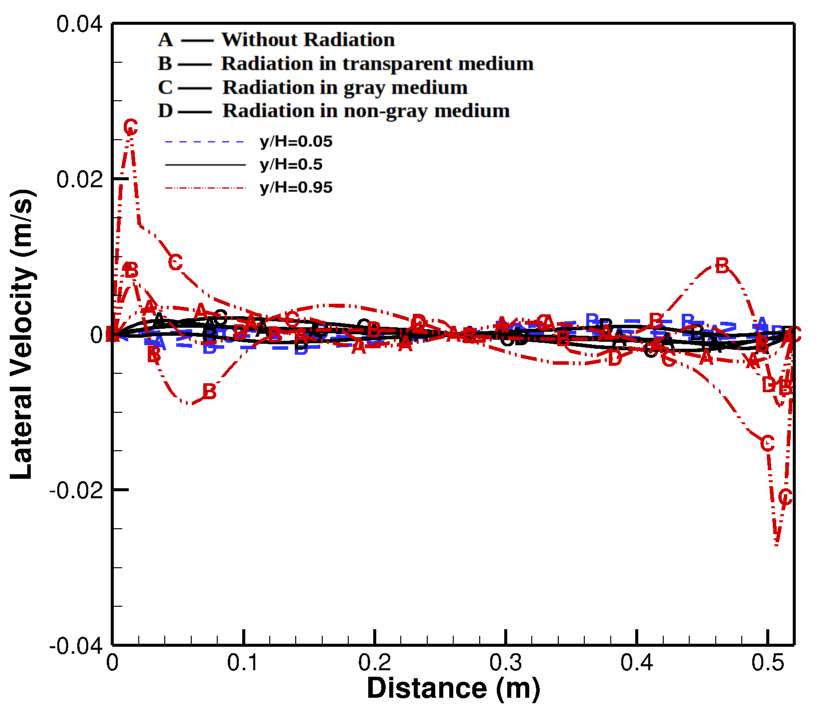}\\
    \mathrm{(a)~~ (near~~ to~~ the~~ cold~~ wall)} & \mathrm{(b)} \\
    \multicolumn{2}{c}{\includegraphics[width=0.5\textwidth]{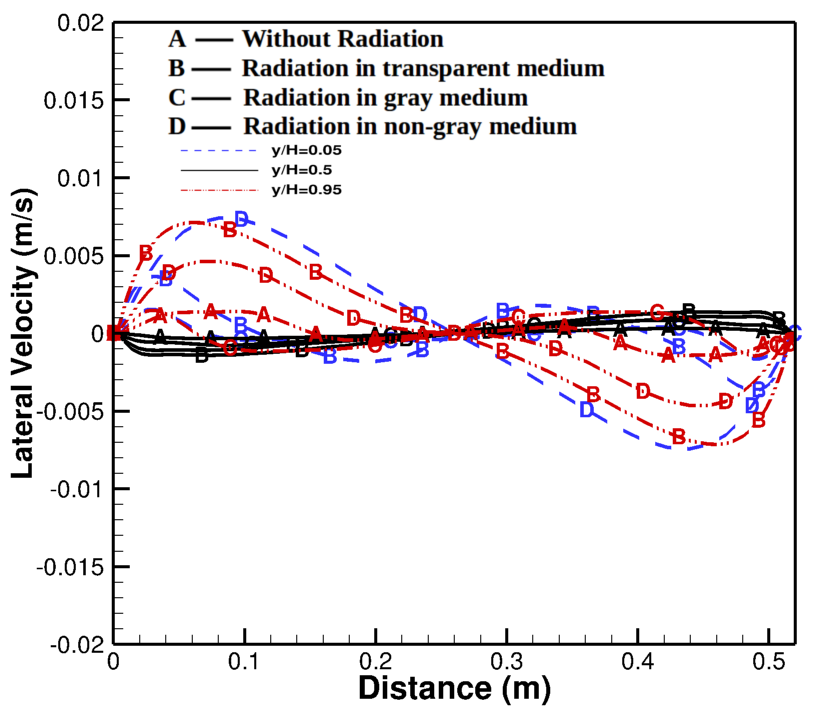}}\\
    \multicolumn{2}{c}{(c)}\\
    \includegraphics[width=0.5\textwidth]{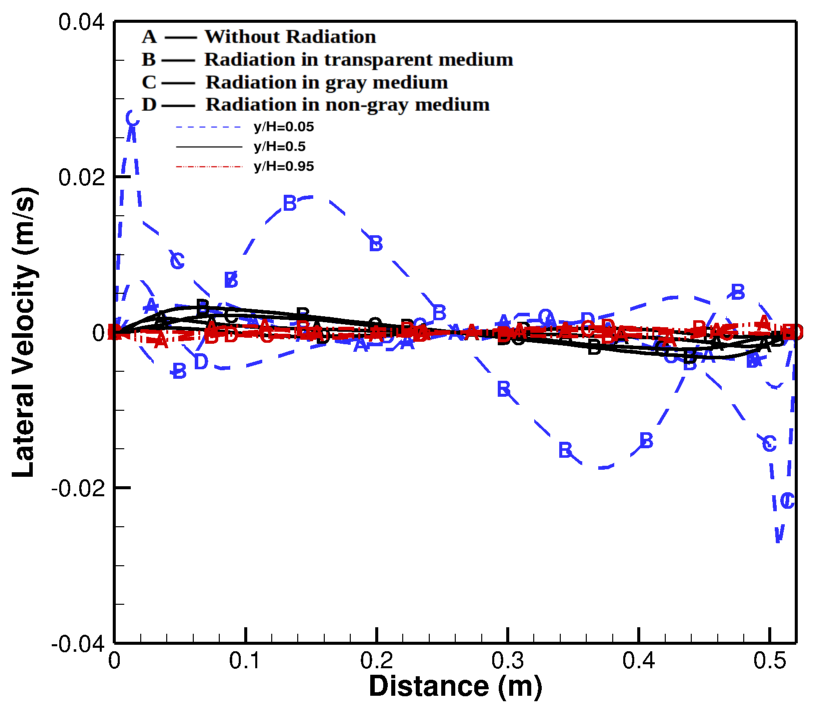} &
    \includegraphics[width=0.5\textwidth]{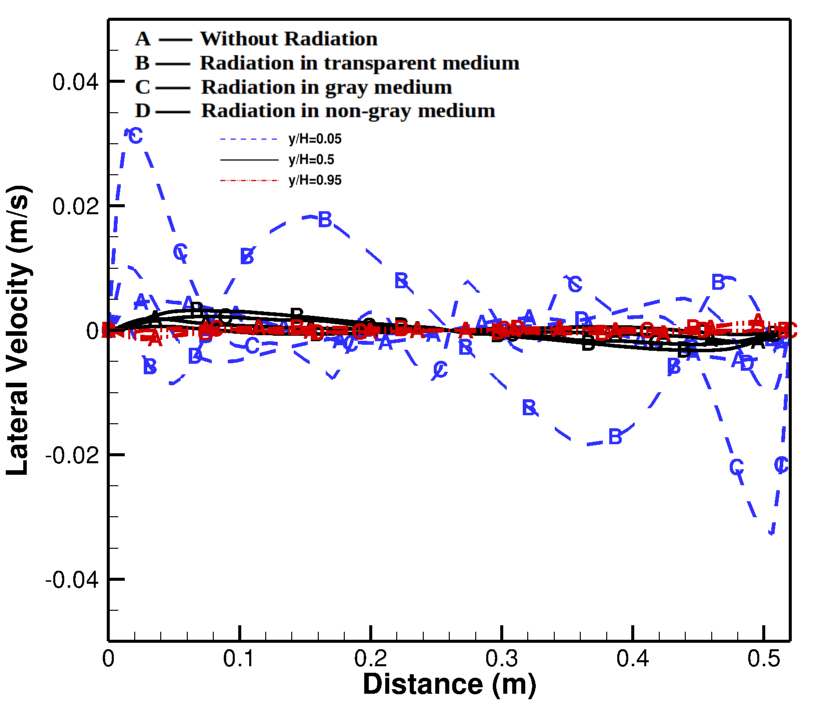}\\
    (d) & (e)\\
\end{array}$
\caption{\label{fig:SqVz}Variation of the lateral velocity ($V_{z}$) along a line at various heights on planes parallel to the isothermal wall at a distance (a) 0.01 m (b) 0.015 m (c) 0.26 m (d) 0.505 m (e) 0.51 m from the cold wall}
\end{figure*}

Figure \ref{fig:SqVz} shows the variation of lateral velocity $(u_{3})$ along a line parallel to the active walls at various heights (a) 0.01 m (b) 0.015 m (c) 0.038 m (d) 0.061 m (e) 0.066 m from the cold wall. The lateral velocity $(u_{3})$ is almost two-order less than the vertical velocity ($u_{2}$) near to the walls all the cases except the case where medium is treated as transparent and gray for radiation, also only at the top and the bottom of the cavity. The variation of lateral velocity in transparent and gray media is wavy in nature and form a wave with variable amplitude in positive and negative direction the top of the cavity near to the cold wall and the bottom of cavity near hot wall. Moreover, these waves are asymmetric about the mid-point of the graph.This wavy nature of the lateral velocity at the top near to the cold wall diminishes as we move towards the center of the cavity. The wavy nature of lateral velocity is also observed in the cases of pure convection and non-gray medium but the magnitude of amplitude is quite small compared to the transparent and gray medium cases. The lateral velocity is nearly zero at mid height of the cavity for all scenarios of radiation modeling.

\begin{figure*}[]
$\begin{array}{cc}
    \includegraphics[width=5cm,height=9cm]{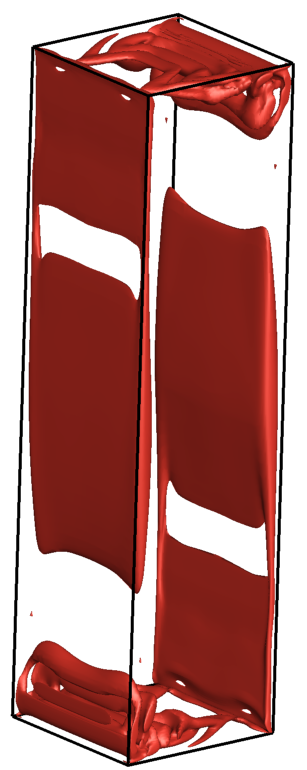} &
    \includegraphics[width=5cm,height=9cm]{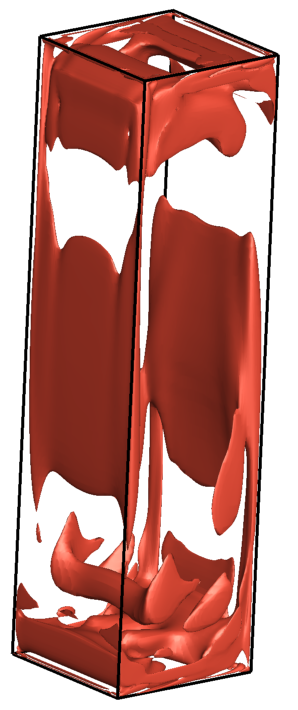}\\
    \mathrm{(A)} & \mathrm{(B)} \\
    \includegraphics[width=5cm,height=9cm]{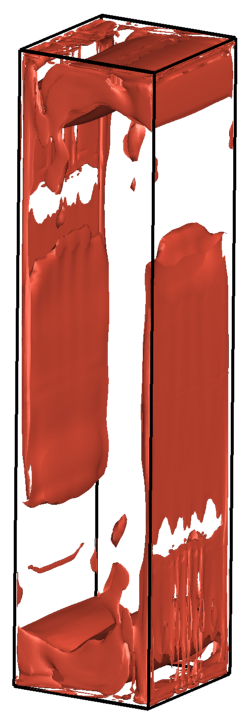} &
    \includegraphics[width=5cm,height=9cm]{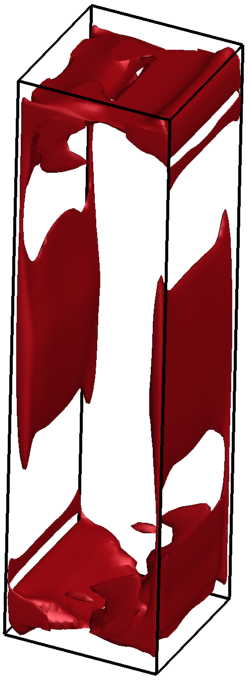}\\
    (C) & (D)\\
\end{array}$
\caption{\label{fig:qSq}Iso-surface of Q criterion for a value of 0.03 for (A) pure convection, and (B) combined convection and radiation in transparent medium, (C) gray medium, (D) non-gray medium}
\end{figure*}

Figure \ref{fig:qSq} depicts the iso-surface of Q criterion for a value of 0.03 (magnitude of vorticity tensor exceeds the rate of strain tensor). The iso-surface for Q criterion changes significantly from case (A) to (D) for this cavity unlike the slender cavity case and this may be direct consequence of radiation heat transfer. The vortices are observed on the active walls and also at the top and the bottom near to the cold and the hot walls, respectively of the cavity in the pure convection case while inclusion of radiation has led to change in the vortice structure and now these vortices are also found most part of the to top and the bottom walls of the cavity. The isosurfaces of Q-criteria are smooth on the active walls for three cases, i.e, pure convection, radiation in transparent and non-gray medium, while fibger structure of vortices are observed for the radiation in the gray medium. The vortices structures are of different shapes on the top and the bottom walls in case of radiation in transparent and gray medium cases, while few structures are found in pure convection case and more in radiation in non-gray medium on top and bottom walls. These changes in the flow structure are the direct results of absorption pattern of radiation energy as radiation is only absorbed on the wall in radiation in transparent medium case and uniformly absorbed in gray medium case and band wise absorption happens in non-gray medium. The radiation induces a three-dimensional flow structure inside the cavity mainly at the top and bottom of the cavity. On comparing case (B) (transparent medium) with case (C) (gray medium), it is observed that the increase in the absorption coefficient leads to the change in the characteristic of vortices at the top and the bottom of the cavity and also on the vertical active walls of the cavity. The nature of vortices in non-gray radiation case is similar pure convection case, but more volume occupied by the vortices in former case. This may be according to "Wein's displacement law", the maximum emission occurs at 9.76 $\mu m$ wavelength (mean temperature of 298 K) in the present study, but, the absorption coefficient is quite low for in this band (Table 2), and also the low emissivity ($\epsilon = 0.35$) on the active walls leads to less effect of radiation on the fluid flow. Another interesting fact to notice that vortices do not appear on the vertical passive walls. This may be due to fact that there is no fluid motion on these walls.

\subsubsection{Heat Transfer Characteristics }

\begin{figure}[t!]  
	\begin{center}
		\includegraphics[width =11cm,height=7cm]{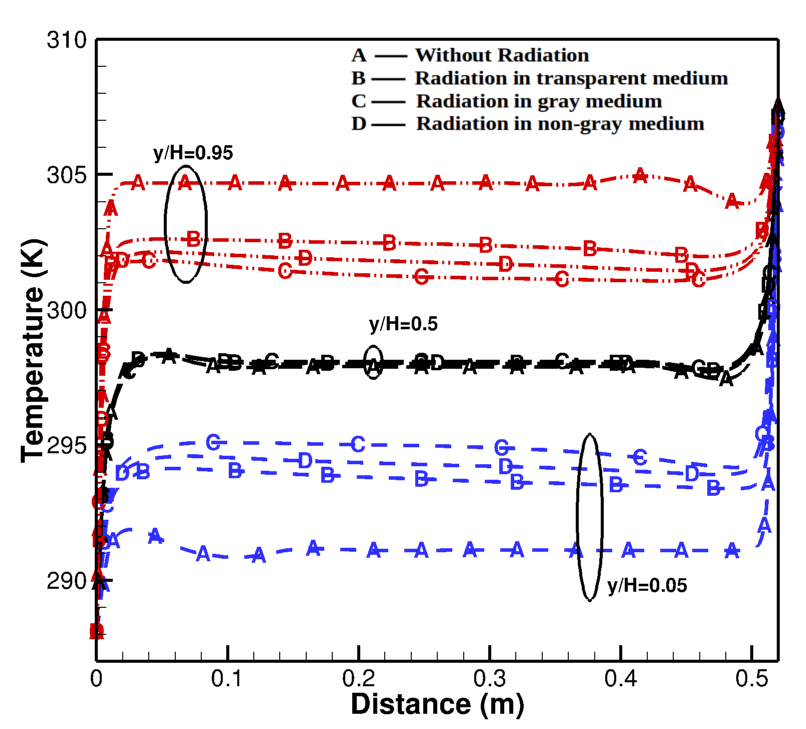}
		\caption{Variation of the temperature along the line at various vertical heights on the mid-plane perpendicular to isothermal walls of the cavity}
		\label{Fig:varTempSq}
	\end{center}
\end{figure}
Figure \ref{Fig:varTempSq} depicts the temperature variation along the line at various vertical heights on the mid plane perpendicular to the active walls of the cavity for the four different scenarios of radiation modeling. The temperature profiles for all the four scenarios are indistinguishable at y/H = 0.5, however, substantial differences appear in all four cases at the top and the bottom of the cavity. The temperature is the highest at the top of the cavity for pure convection case, whereas, it is the lowest at the bottom of the cavity. The average temperature at the top of the cavity for the four different scenarios in ascending order is as radiation in gray medium, non-gray medium, transparent medium and without radiation (pure convection case). While, reverse trend is observed at the bottom of the cavity, i.e., the trend in ascending order is as; without radiation (pure convection case), radiation in transparent medium, non-gray medium and gray medium. The average temperature difference between pure convection and radiation in gray medium cases is almost 3 K at y/H = 0.95, while this difference increases to 4 K at y/H = 0.05, which are quite significant as operating temperature inside of the cavity is around room temperature. The temperature variations for any scenarios of are only seen in a small distance of 0.04 m from the active walls at mid-height, this reveals the fact that heat transfer only happens with in small distance from the active wall and thisr does not happen within core of the cavity, i.e., the fluid in the core itself imposing a adiabatic condition.
\begin{table}[t!]
	\begin{center}
	\caption{\label{avaTempSqtab}Average temperature inside the square-base tall cavity for four scenarios of radiation modeling}
		\begin{tabular}{|c|c|c|}
			\hline
			S.No. & Case     & Average temperature \\ \hline
			1    & Case (A) & 297.89 K        \\ \hline
			2    & Case (B) & 298.01 K        \\ \hline
			3    & Case (C) &  298.05K        \\ \hline
			4    & Case (D) &  298.02K        \\ \hline
		\end{tabular}
	\end{center}
\end{table}

\begin{figure*}[]
$\begin{array}{cc}
    \includegraphics[width=5cm,height=9cm]{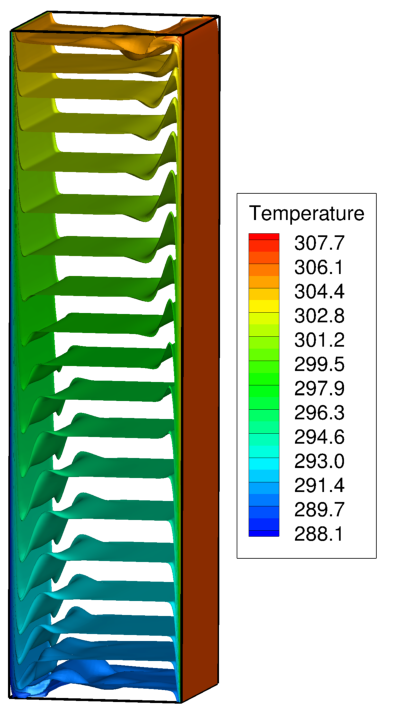} &
    \includegraphics[width=5cm,height=9cm]{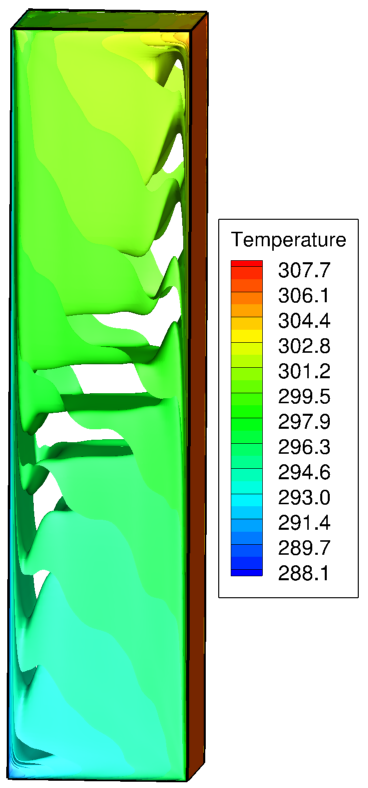}\\
    \mathrm{(A)} & \mathrm{(B)} \\
    \includegraphics[width=5cm,height=9cm]{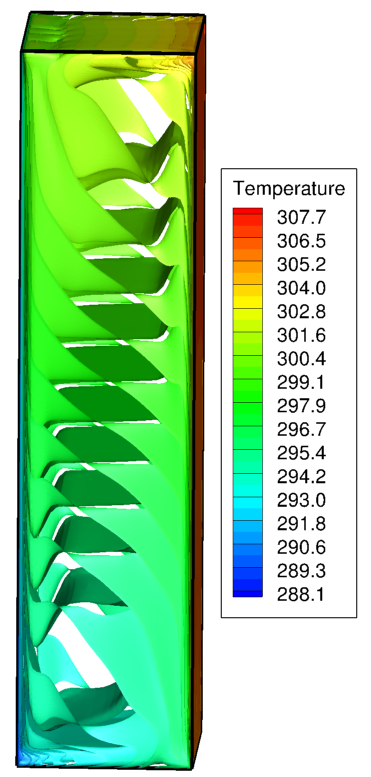} &
    \includegraphics[width=5cm,height=9cm]{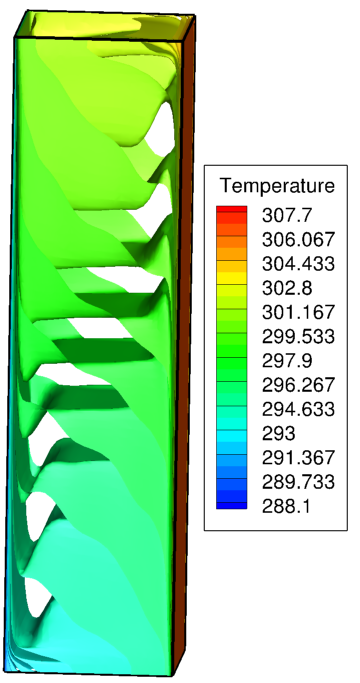}\\
    (C) & (D)\\
\end{array}$
\caption{\label{fig:tempSq}Isosurface the of temperature (K) for (A) pure convection, and (B) combined convection and radiation in transparent medium, (C) gray medium, (D) non-gray medium}
\end{figure*}

\begin{figure*}[]
$\begin{array}{cc}
    \includegraphics[width=5cm,height=9cm]{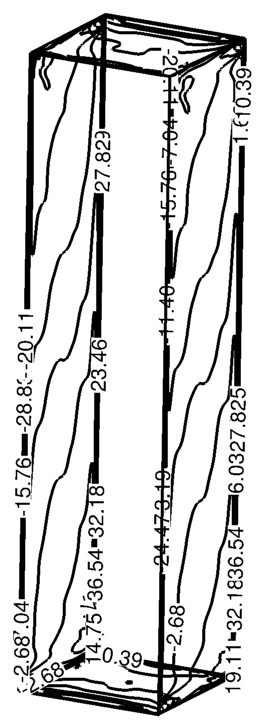} &
    \includegraphics[width=5cm,height=9cm]{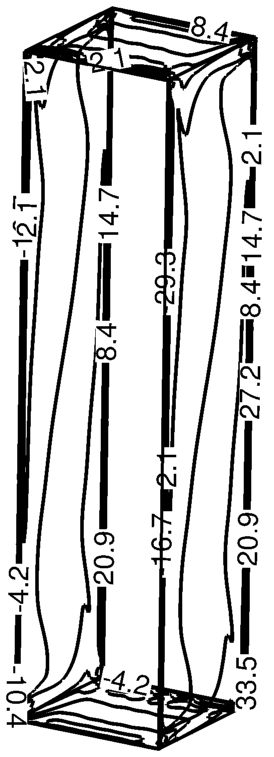}\\
    \mathrm{(B)} & \mathrm{(C)} \\
    \multicolumn{2}{c}{\includegraphics[width=5cm,height=9cm]{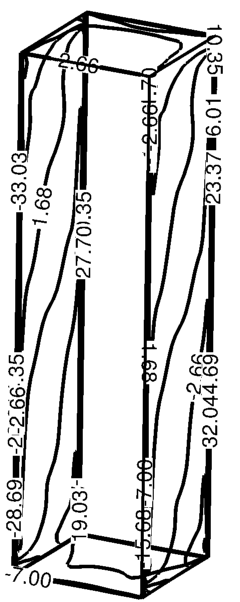}}\\
   \multicolumn{2}{c}{(D)}\\
\end{array}$
\caption{\label{fig:radHeat}Distribution of the radiative heat flux $(W/m^{2})$ on the adiabatic walls for (B) combined convection and radiation in transparent medium, (C) gray medium, and (D) non-gray medium}
\end{figure*}

\begin{figure*}[]
$\begin{array}{cc}
    \includegraphics[width=5cm,height=9cm]{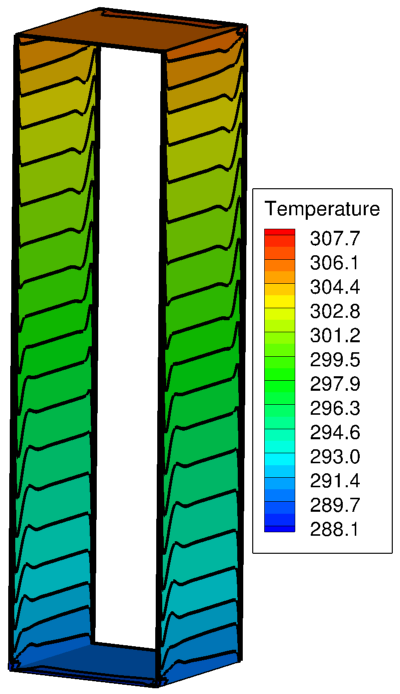} &
    \includegraphics[width=5cm,height=9cm]{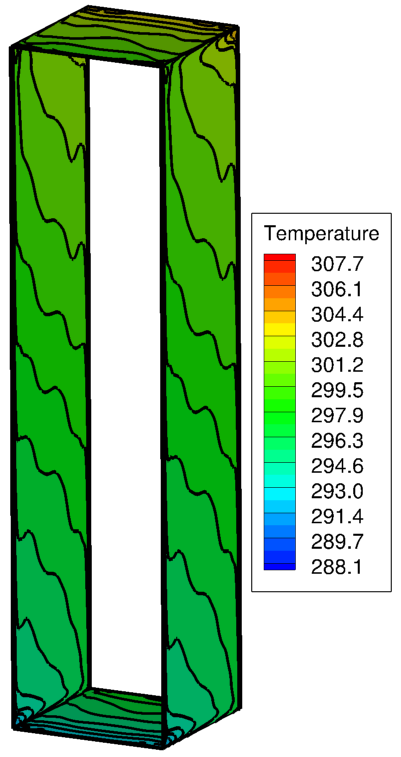}\\
    \mathrm{(A)} & \mathrm{(B)} \\
    \includegraphics[width=5cm,height=9cm]{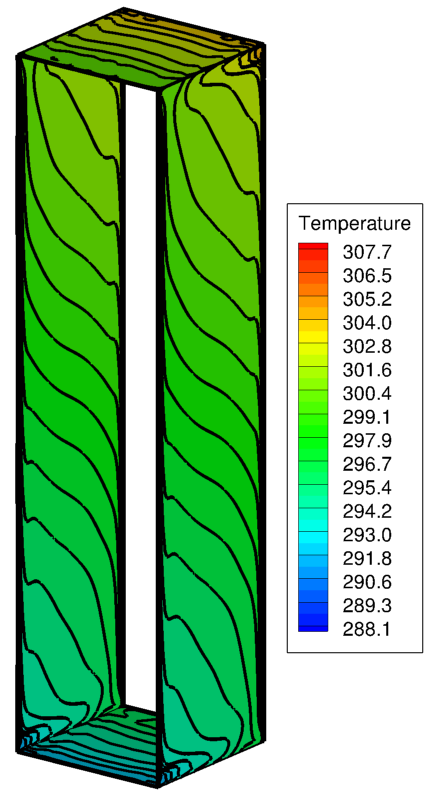} &
    \includegraphics[width=5cm,height=9cm]{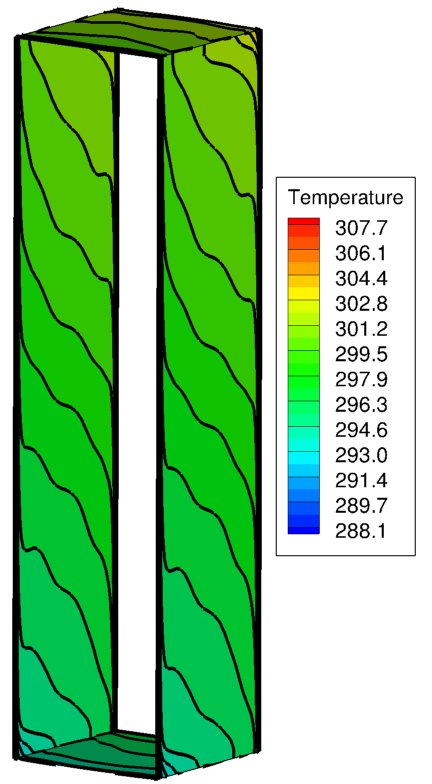}\\
    (C) & (D)\\
\end{array}$
\caption{\label{fig:tempAd}Distribution of the temperature (K) on the adiabatic walls (A) pure convection, and (B) combined convection and radiation in transparent medium, (C) gray medium, (D) non-gray medium}
\end{figure*}

Figure \ref{fig:tempSq} shows the  iso-surface of temperature inside the cavity for (A) pure convection, and (B) combined convection and radiation in transparent, (C) gray, and (D) non-gray medium. A thermal stratification is observed inside the cavity in the pure convection case, whereas this stratification gets disturbed with the inclusion of the radiation mode of heat transfer. The temperature iso-surfaces are folded along the passive walls on the cavity with the inclusion of the radiation, nevertheless, the thermal stratification is observed inside the cavity. This is due to the fact that the temperature gradient on the passive walls is no longer perpendicular to the walls and decided by both the conduction and radiative heat fluxes, thus, the temperature field inside the cavity. The nature of radiative flux changes from negative to positive along the diagonal line on the vertical passive walls as shown in Fig. \ref{fig:radHeat} and opposite to this should be the nature of the conductive heat flux as total heat flux is zero on the passive walls. This causes the change in gradient direction along the diagonal line and therefore, the folding of the temperature isotherms also happens in the upward and the downward directions.  The folding of the temperature iso-surface is reduced little in case of gray medium but gets further enhanced for non-gray medium. A little thermal stratification formation is seen at the corners (top left and bottom right) for the gray medium case, while it reduces in the case of non-gray medium. Overall, a complex temperature structures appears in the cases where radiation is involved.

\begin{figure*}[]
$\begin{array}{cc}
    \includegraphics[width=5cm,height=9cm]{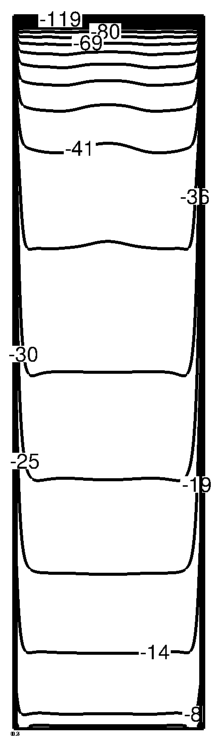} &
    \includegraphics[width=5cm,height=9cm]{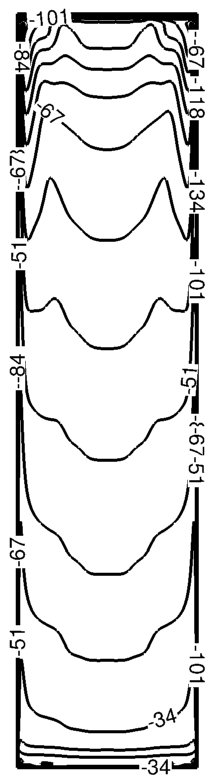}\\
    \mathrm{(A)} & \mathrm{(B)} \\
    \includegraphics[width=5cm,height=9cm]{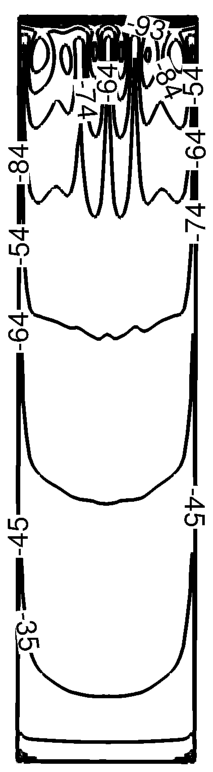} &
    \includegraphics[width=5cm,height=9cm]{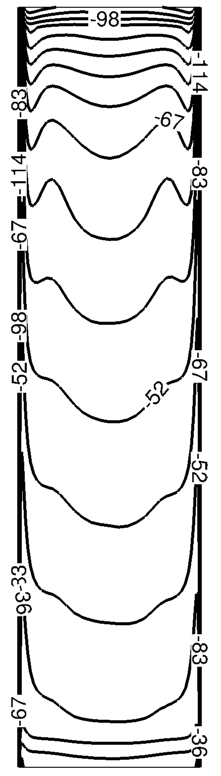}\\
    (C) & (D)\\
\end{array}$
\caption{\label{fig:totNussSq}Total Nusselt number contours on the cold wall (288.1 K) for (A) pure convection, and (B) combined convection and radiation in transparent medium, (C) gray medium, (D) non-gray medium}
\end{figure*}

\begin{figure*}[]
$\begin{array}{cc}
    \includegraphics[width=5cm,height=9cm]{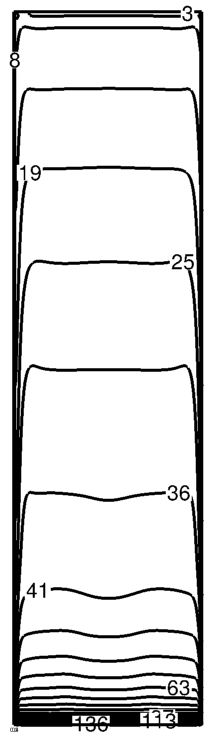} &
    \includegraphics[width=5cm,height=9cm]{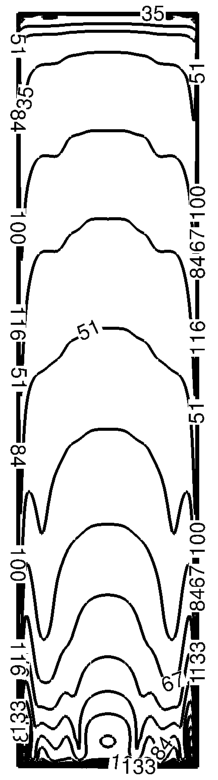}\\
    \mathrm{(A)} & \mathrm{(B)} \\
    \includegraphics[width=5cm,height=9cm]{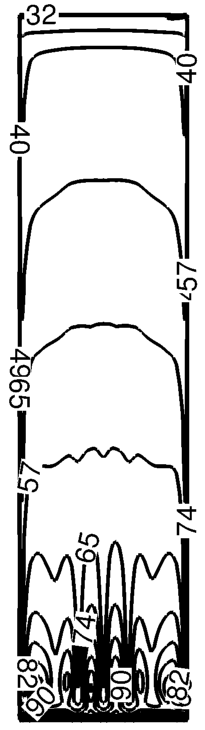} &
    \includegraphics[width=5cm,height=9cm]{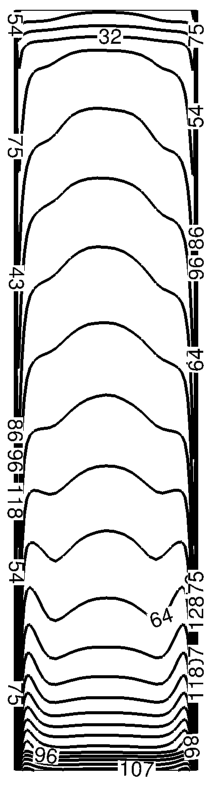}\\
    (C) & (D)\\
\end{array}$
\caption{\label{fig:totNussHot}Total Nusselt number contours on the hot wall (307.7 K) for (A) pure convection, and (B) combined convection and radiation in transparent medium, (C) gray medium, (D) non-gray medium}
\end{figure*}

The complex structure of the temperature iso-surfaces inside the cavity with inclusion of the radiation have been formed, it is interesting to know the temperature profile on these passive walls. The temperature distributions on passive walls are shown in Fig. \ref{fig:tempAd} for (A) pure convection, and (B) combined convection and radiation in transparent, (C) gray, (D) non-gray medium. The isolines of temperature are almost horizontal on the most portion of the vertical passive walls, however, little variations are observed near to the active walls for pure convection case. Whereas, these isotherm lines are slanted on the vertical passive walls and almost parallel to the active walls on the horizontal passive walls for the radiation cases. It can be also observed that isothermal lines are little wavy for radiation in transparent medium while this gets smoothen for radiation in gray medium. The nature of waviness reduces for radiation in non-gray medium case. Also, the angle of these slanted isothermal lines on the vertical adiabatic walls has increased for gray medium and again decreases for the non-gray medium compared to the transparent medium. Nevertheless, the temperature within the cavity remains remains in the limit of the active walls conditions, i.e., 288.1 K to 307.7 K for all cases.

Inspite of all these variations in the temperature distribution, there is no major change in the volume average temperature inside the square-base tall cavity as can be seen in Table \ref{avaTempSqtab}. This could be due to fact that emissions are small in those band where absorption coefficient is high, thus overall small absorption of radiation energy inside the cavity, also since most portion of the fluid in core remains at constant temperature, thus, causes absorption and emission equal, thus, this lead to no further increase in volume average temperature inside the cavity.

The total Nusselt number variation on the cold wall for (A) pure convection, and (B) combined convection and radiation in transparent medium, (C) gray medium, (D) non-gray medium are depicted in Fig. \ref{fig:totNussSq}. The Nusselt number vary considerably from the top to the bottom for all four cases considered for the study. The minimum and maximum Nusselt numbers are 1.17 and 136.9, respectively for pure convection, furthermore, these values are 16.9, and 150.5 for transparent medium, 15.8 and 156.67 for gray medium and 20.2 to 143.97 for non-gray medium cases. The interesting fact to notice is that Nusselt number changes significantly with the inclusion of radiation. The minimum Nusselt number decreases and the maximum Nusselt number increases for gray medium case compared to transparent medium case while reverse happens for non-gray medium case. Furthermore, the difference in the maximum and the minimum Nusselt number remains same for all scenarios. The closed contours of Nusselt number on the top of the wall for case (C) is attributed to the flow pattern as we have seen in Fig \ref{fig:SqVz} existed at the top of the cavity, where nature of lateral velocity is wavy and existence of the different structures of eddies is found in Fig. \ref{fig:qSq}. Also, the centres of these close contours of the Nusselt number on the top of the cold wall coincides with the high amplitude of the lateral velocity. The similar but the opposite trend for the Nusselt number is observed on the hot wall as shown in Fig. \ref{fig:totNussHot}. 

Table \ref{heatFluxSq} represents the area average wall heat flux ($W/m^{2}$) on the active and passive walls of the square-base tall cavity for the cases (A), (B), (C), and (D). The conductive heat flux (CHF) and radiative heat flux (RHF) remain almost same for all scenarios, but there are less than the slender cavity case. The conductive heat flux is zero on adiabatic walls for pure convection case, however this is not true for the combined modes of heat transfer, because the sum of both fluxes is zero on the passive walls. Thus, conductive flux is equal and opposite to the radiative flux on the passive walls. These fluxes are high on horizontal passive  walls, while it is very small on vertical passive walls.

\begin{table}[b!]
	\centering
	\addtolength{\tabcolsep}{-4pt}
	\caption{Heat flux ($W/m^{2}$) report of square-base tall cavity for (a) pure convection, and (B) combined convection and radiation in transparent medium, (C) gray medium, (D) non-gray medium}
	\label{heatFluxSq}
	\begin{tabular}{|c|l|c|c|c|c|c|c|c|}
		\hline
		\multicolumn{2}{|c|}{\multirow{2}{*}{Walls}} & Case (A) & \multicolumn{2}{c|}{Case (B)} & \multicolumn{2}{c|}{Case (C)} & \multicolumn{2}{c|}{Case (D)} \\ \cline{3-9} 
		\multicolumn{2}{|c|}{}                       & CHF   & CHF         & RHF         & CHF         & RHF         & CHF         & RHF         \\ \hline
		\multicolumn{2}{|c|}{Isothermal Cold }                   & -30.025  & -29.946       & -21.541       &  30.85      & -20.03       & -30.409       & -21.107       \\ \hline
		\multicolumn{2}{|c|}{Isothermal Hot}                  & 30.025   & 29.946       & 21.541        & 29.624        & 21.116        & 29.465        & 22.051        \\ \hline
		\multicolumn{2}{|c|}{Adiabatic Top }                    & 0        & -6.777       & 6.777        & -4.473      & 4.473        & -4.263       & 4.263        \\ \hline
		\multicolumn{2}{|c|}{Adiabatic Bottom }                 & 0        & 7.248        & -7.248       & 4.687        & -4.687      & 4.714        & -4.714       \\ \hline
		\multicolumn{2}{|c|}{Adiabatic Front }                  & 0        & 0.387         & -0.387        & 0.092         & -0.092        & 0.228        & -0.228       \\ \hline
		\multicolumn{2}{|c|}{Adiabatic Back }                  & 0        & 0.387         & -0.387        & 0.119         & -0.119        & 0.228        & -0.228         \\ \hline
	\end{tabular}
\end{table}

\section{Conclusions}

Two three-dimensional differentially heated tall cavities whose height and width are 2.18 m and 0.52 m, respectively, while distance between active walls are 0.076 m (slender cavity) and 0.52 m, (square base), respectively, are considered to investigate the effects of non-gray/gray radiation modeling on the natural convection. The active walls are maintained at temperature 288.1 K and 307.7 K for both the cavities and rest are the passive walls. The thermophysical quantities are kept same and Boussinesq approximation has been used to model the buoyancy for both the cavities. Four different scenarios of radiation modeling, i.e., (A) pure convection or without radiation, and (B) combined convection and radiation in transparent medium, (C) gray medium, (D) non-gray medium have been considered. The commercial CFD package ANSYS Fluent has been used for the analysis of fluid flow and heat transfer in the natural convection and bands model of the radiation for the non-gray analysis of radiation. First, the computational results for four scenarios of radiation modeling of the slender cavity have been compared with the experimental results \cite{betts2000experiments}, and comprehensive results of slender cavity were presented. The investigation is further extended for the study of the dynamics of fluid flow and heat transfer for square-base cavity have been analyzed. Based on the above studies, the following conclusions are drawn
\begin{enumerate}

	\item The temperature profiles at the top and the bottom of the cavities show a difference in all four scenarios of the radiation modeling, however, no difference in the temperature profile is seen in the middle of the slender cavity, and hardly any difference is found in the velocity profile for all four scenarios of the radiation modeling.
	
	\item The maximum and the minimum temperatures are found in pure  convection case and radiation in gray medium, respectively, at the top of both the cavities and the difference in the temperatures profile is approximately 2 K and 3 K for slender and square base cavities, respectively. The temperature profile at the bottom of the cavities is in reverse order of the temperature profile at the top, however, the difference in the temperature profile at the bottom of the cavity is 3 K and 4 K for the slender and the square base cavities, respectively.

	\item The lateral velocity is one order less to the vertical velocity at the top and the bottom of the slender and the square base cavity, whereas, this is two order less at mid height of the cavity. The behaviour of lateral velocity is wavy in nature in the square-base cavity for the transparent and gray medium cases.
	
	\item The conductive and the radiative flux reports of slender cavity match very well with experimental results and are of same order of magnitude. These fluxes are also found to be almost same for the square cavity.
	
	\item The flow characteristics inside the slender cavity remains unaffected in all scenarios of radiation modeling, however, the difference arises in square base cavity. The radiation leads to enhancement in the formation of the vortices on the top and the bottom of the cavity. The transparent and the gray medium produce many different structures of eddies. However, no eddies are formed on the vertical passive walls of the cavities.
	
	\item The fluid flow and heat transfer are only happening in a very small distance from the active walls and from the top and the bottom of the square base cavity. The flow in the core the cavity is at rest and the fluid remains isothermal.
	
	\item The different scenarios of radiation modeling only show effect on the fluid flow and heat transfer at the top and the bottom of the cavity and there remain unaffected elsewhere  in the cavity.
	
\end{enumerate}

%\section*{Acknowledgements}
%\acknowledgements
%The authors greatly acknowledge the financial support provided by Science and Engineering Research Board (SERB) (Statutory Body of the Government of India) via Grant.No:ECR/2015/000327 to carry out the present work.

\section*{Declaration of interest}
The authors declare that they have no known financial interests or personal relationships that could have appeared to influence the work reported in this paper.

% \usepackage{multirow}

%\section{Bibliography styles}

%\section*{References}

\bibliography{nonGrayTall}

\end{document}